\documentclass[envcountsame]{tlp}

\usepackage{amssymb}
\usepackage{amsmath}
\usepackage{subfigure}
\usepackage{epsfig}
\usepackage{graphics}
\usepackage{url}

\newcommand{\incase}{:\!\!-\;}

\newcounter{polycons2ctr}
\newcounter{auxctr}
\newcounter{polycons1ctr}
\newcounter{ex:der-lastsymconsctr}

\newtheorem{definition}{Definition}
\newtheorem{example}{Example}
\newtheorem{theorem}{Theorem}

\newtheorem{corollary}{Corollary}

\newtheorem{proposition}{Proposition}

\newtheorem{procedure}{Procedure}

\submitted {18 February 2008}
\revised {30 November 2009}
\accepted{11 December 2009}

\author[Manh Thang Nguyen et al.]{
    MANH THANG NGUYEN \\
	Deceased on June 3, 2009
    \and DANNY DE SCHREYE \\
	Department of Computer Science, K.~U.\ Leuven \\ Celestijnenlaan 200A, B-3001 Heverlee,
    Belgium \\
    Danny.DeSchreye@cs.kuleuven.ac.be
    \and J\"{U}RGEN GIESL \\
	LuFG Informatik 2, RWTH Aachen \\ Ahornstr.\ 55,  D-52074 Aachen, Germany \\
    giesl@informatik.rwth-aachen.de
    \and PETER SCHNEIDER-KAMP  \\
	Dept. of Mathematics and Computer Science, U.\ Southern Denmark \\ Campusvej 55, DK-5230 Odense M,
    Denmark \\
    petersk@imada.sdu.dk
        }

\title[Polynomial Interpretations for Termination Analysis of Logic Programs]{\textsf{Polytool}: Polynomial Interpretations as a Basis for Termination Analysis of Logic Programs}
\begin{document}
\maketitle
\begin{abstract}
%
Our goal is to study the feasibility of
porting termination analysis techniques developed for one programming paradigm to 
another paradigm. 
In this paper, we show how to adapt
termination analysis techniques based on polynomial interpretations - very well
known in the context of term rewrite systems (TRSs) -
to obtain new (non-transformational) 
termination analysis techniques for definite logic programs (LPs). 
This leads to an approach that can be seen as a
direct generalization of the traditional techniques in termination analysis of LPs, where
linear norms and level mappings are used. Our extension generalizes these to
arbitrary polynomials. We extend a number of standard concepts and results on termination
analysis to the context of polynomial interpretations. We also propose a constraint-based
approach for automatically generating polynomial interpretations that satisfy the
termination conditions. Based on this approach, we implemented a new tool, called
\textsf{Polytool}, for automatic termination analysis of LPs.
\\
\\
\end{abstract}


\section{Introduction}\label{sec:intro}
Termination analysis plays an important role in the study of program correctness. A
termination proof is mostly based on a mapping from computational states to some
well-founded ordered set. Termination is guaranteed if the mapped values of the
encountered states during a computation, under this mapping, decrease w.r.t.\ the
order.

For LPs, termination analysis is done by mapping terms and atoms to a
well-founded set of natural numbers by means of norms and level mappings. Proving
termination is based on the search for a suitable norm and level mapping such
that
%
the resulting predicate calls decrease under the
mapping. 

Until now, most termination techniques for LPs are based on the use of
linear
norms and linear level mappings,
which measure the size of each term or atom as a linear combination of the sizes of its
sub-terms. For example, the \textsf{Hasta-La-Vista} system
\cite{SerebrenikandDeSchreye03} 
infers one specific linear norm and linear level mapping. In the context of numerical
computations, it includes a refinement on this, based on a case analysis. 
The tool \textsf{cTI}
\cite{MesnardBagnara05} uses a concrete linear norm. The analyzers \textsf{TermiLog}
\cite{lindenstrauss97,Termilog} and \textsf{TerminWeb} \cite{Codishetal99,terminWeb02} use
a combination of several linear norms to obtain an
approximation of the program and then infer linear level mappings for termination
analysis of the approximated program. However, the
restriction to linear norms and level mappings limits the power of termination analysis
considerably. 
To
illustrate this point, consider the following example, \emph{der}, that formulates rules
for computing the repeated derivative of a function in some variable
$u$. This example from \cite{DeSchreyeSerebrenik01,NaomiWST97} is inspired by a
similar term rewriting example 
from \cite{Dershowitz95}.

\begin{example}[der]\label{exam:der}
\begin{align}
  &d(\mathit{der}(u),1). \label{der1} \\
  &d(\mathit{der}(X+Y),\mathit{DX}+\mathit{DY}) \incase d(\mathit{der}(X),\mathit{DX}),
d(\mathit{der}(Y),\mathit{DY}). \label{der2} \\ 
  &d(\mathit{der}(X*Y),X*\mathit{DY}+Y*\mathit{DX}) \incase
d(\mathit{der}(X),\mathit{DX}),d(\mathit{der}(Y),\mathit{DY}). \label{der3} \\ 
  &d(\mathit{der}(\mathit{der}(X)),\mathit{DDX}) \incase d(\mathit{der}(X),\mathit{DX}),
d(\mathit{der}(\mathit{DX}),\mathit{DDX}). \label{der4} 
\end{align}
We are interested in proving termination of this program w.r.t.\ the set of queries $S=\{\,
d(t_1,t_2) \mid t_1$ is a ground term and $t_2$ is an arbitrary term\}.  So
the set of queries is specified by a \emph{mode} that considers
the first argument of $d$ as an input argument and the second as an output.

As
shown in \cite{NaomiWST97,MNTDannyd05}, the termination proof is impossible when using
a linear norm and a linear level mapping. Indeed,
it turns out that all existing non-transformational termination analyzers for LPs
mentioned above fail to prove termination of this example.
%
{\hfill{$\square$}}
\end{example}

In this paper, we propose a general framework for termination proofs of LPs based on
polynomial interpretations. Using polynomial interpretations as a basis for ordering
terms in TRSs was first introduced by Lankford in
\cite{Lankford79}. It is currently one of the best known and most widely used techniques
in TRS termination analysis.

We develop the approach within an LP context. 
Classical approaches in LP termination use 
interpretations that map to
natural numbers (using linear polynomial functions). In contrast, we will use interpretations that
map to polynomials (using arbitrary polynomial functions). To adapt the
classical LP approaches to polynomial interpretations, we use the concepts
of ``abstract norm'' and ``abstract level mapping''
\cite{Verschaetse&DeSchreye91}.
We show that with our new approach, one can also prove termination of programs
like Example \ref{exam:der}.

We also developed an automated tool (\textsf{Polytool}) for termination analysis based on
our approach \cite{Nguyen&DeSchreye06}. 
We embedded this within the constraint-based approach developed in
\cite{Decorteetal98} and combined it with the non-linear Diophantine constraint solver
developed by  Fuhs et al.\ \cite{Fuhsc07} (implemented in the \textsf{AProVE} system
\cite{Giesletal06}) to provide a completely automated system.

The paper is organized as follows. In the next section, we present some preliminaries. In
Section \ref{sec:interpretation}, we introduce the notion of polynomial interpretations
in logic programming and show how this approach can be used to prove termination.
In Section \ref{sec:automation}, we discuss the automation of the approach. In
Section \ref{sec:experiment}, we provide and discuss the results of our experimental
evaluation. We end with a conclusion in Section \ref{conclusion}.

%
\section{Preliminaries}\label{sec:preliminaries}

After
introducing the basic terminology of LPs in Section
\ref{subsec:notation}, we recapitulate the concepts of \emph{norms} and \emph{level
mappings} in Section \ref{subsec:norm&levelmapping} and explain their use for termination
proofs in Section \ref{subsec:conditions}.

\subsection{Notations and Terminology}\label{subsec:notation}

We assume familiarity with LP concepts and with the main results of logic programming
\cite{Apt90,Lloyd87}. In the following, $P$ denotes a definite logic program.
We use $\mathit{Var}_P$, $\mathit{Fun}_P$, and $\mathit{Pred}_P$ to denote the sets of
variables, function, and predicate symbols of $P$. Given an atom $A$, $\mathit{rel}(A)$
denotes the predicate occurring in $A$. Let $p$, $q$ be predicates occurring in the
program $P$. We say that $p$ \emph{refers to} $q$ if there is a clause in $P$ such that
$p$ is in its head and $q$ is in its body. We say that $p$ \emph{depends on} $q$ if
$(p,q)$ is in the transitive closure of the relation ``refers to''. If $p$ depends on $q$
and vice versa, $p$ and $q$ are called \emph{mutually recursive}, denoted by $p
\backsimeq q$. A clause in $P$ with a predicate $p$ in its head and a predicate $q$ in
its body, such that $p$ and $q$ are mutually recursive, is called a 
\emph{(mutually) recursive clause}. Within such a recursive clause, the body-atoms 
with predicate symbol $q$ are called \emph{(mutually) recursive atoms}. 
Let $\mathit{Term}_P$ and $\mathit{Atom}_P$ denote, respectively, the sets
of all terms and atoms that can be constructed from $P$. 

In this paper, we focus our attention on definite logic programs and SLD-derivations
where the left-to-right selection rule is used. Such derivations are referred to as
LD-derivations; the corresponding derivation tree is called \emph{LD-tree}. We say that a
query $Q$ \emph{LD-terminates} for a program $P$, if the LD-tree for $(P,Q)$ is finite
(left-termination \cite{Lloyd87}). In the following, we usually speak of ``termination''
instead of ``LD-termination'' or ``left-termination''.

\subsection{Norms and Level Mappings}\label{subsec:norm&levelmapping}

The concepts of \emph{norm} and \emph{level mapping} are central in termination analysis
of logic programs.

\begin{definition}[norm, level mapping]\label{def:norm&levelmapping}
A \emph{norm} is a mapping ${\parallel}.{\parallel}: \mathit{Term}_P \rightarrow
\mathbb{N}$. A \emph{level-mapping} is a mapping ${\mid}.{\mid}: \mathit{Atom}_P
\rightarrow \mathbb{N}$.
\end{definition}

Several examples of norms can be found in the literature \cite{Bossietal91}. One of the
most commonly used norms is the \emph{list-length norm}
${\parallel}.{\parallel}_\ell$ which maps lists to their lengths 
and any other term to 0. Another frequently used norm is the \emph{term-size} norm
${\parallel}.{\parallel}_\tau$
which counts
the number of function symbols in a term. Both of them belong to a class of norms called
linear norms which is defined as follows.

\begin{definition}[linear norm and level mapping \cite{Serebrenik03}] \label{def:linearnorm}
A norm ${\parallel}.{\parallel}$ is a \emph{linear} norm if it is recursively defined by
means of the following schema:
\begin{enumerate}
\item[-]${\parallel}X{\parallel} = 0$ for any variable $X$,
\item[-]${\parallel}f(t_1,\ldots,t_n){\parallel} =
f_0 +\sum_{i=1}^{n}f_i{\parallel}t_i{\parallel}$ where $f_i \in \mathbb{N}$ and $n \ge
0$.
\end{enumerate}
Similarly, a level mapping ${\mid}.{\mid}$ is a \emph{linear} level mapping if it is
defined by means of the following schema:
\begin{enumerate}
\item[-]${\mid}p(t_1,\ldots,t_n){\mid} =
p_0 +\sum_{i=1}^{n}p_i{\parallel}t_i{\parallel}$ where $p_i \in \mathbb{N}$ and $n \ge 0$.
\end{enumerate}
\end{definition}
\subsection{Conditions for Termination w.r.t.\ General Orders}\label{subsec:conditions}

A
\emph{quasi-order} on a set $S$ is a reflexive and transitive binary relation
$\succsim$ defined on elements of $S$. We define the \emph{associated equivalence
relation} $\approx$ as $s \approx t$ if and only if $s \succsim t$ and $t \succsim
s$.
A \emph{well-founded order} on $S$ is a transitive relation $\succ$ where there is no
infinite sequence $s_0 \succ s_1 \succ \ldots$ with $s_i \in S$. A \emph{reduction pair}
$(\succsim,\succ)$ consists of a quasi-order $\succsim$ and a well-founded order $\succ$
that are \emph{compatible} (i.e., $t_1 \succsim t_2 \succ t_3$ implies $t_1 \succ
t_3$).
%
%
%
We also need the following notion of a call set.

\begin{definition}[call set]\label{def:callset}
Let $\mathit{P}$ be a program and $\mathit{S}$ be a set of atomic queries. The \emph{call
set}, $\mathit{Call}(P,S)$, is the set of all atoms $\mathit{A}$, such that a variant of
$\mathit{A}$ is the selected atom in some derivation for $\mathit{(P,Q)}$, for some $Q
\in S$.
\end{definition}

Most often, one regards infinite sets $\mathit{S}$ of queries. For instance,
this is the case in Example 1. As in Example 1, $\mathit{S}$ is then specified in 
terms of modes or types. As a consequence, in an automated approach, a safe over-approximation
of $\mathit{Call}(P,S)$ needs to be computed, using a mode or a type inference
technique (e.g.,
\cite{Bruynoogheetal05,GallagherHB05,HeatonACK00,Janssensetal92}).

%
%
%
%
In order to obtain a termination criterion that is
suitable for automation, one usually estimates the effect of the atoms in the
bodies of clauses
by suitable \emph{interargument relations}. This notion can be defined for arbitrary
reduction pairs.

\begin{definition}[interargument relation
\cite{DeSchreyeSerebrenik01}]
\label{def:interargument} Let $P$ be a program, $p$ be a
predicate in $P$, and $(\succsim,\succ)$ be a reduction pair on $\mathit{Term}_P$. An
\emph{interargument relation} for $p$ in $P$ w.r.t.\ $(\succsim,\succ)$ is a relation
$R_p$ with the same arity as p:
$R_p=\{p(t_1,\ldots,t_n) \mid 
 t_i \in \mathit{Term}_P \mbox{ for all $1 \leq i \leq n$, and }
\varphi_p(t_1,\ldots,t_n)\}$, where:
\begin{enumerate}
\item[-] $\varphi_p(t_1,\ldots,t_n)$ is a boolean expression (in terms of
disjunction, conjunction, and negation) of inequalities $s \succsim s'$ or $s \succ s'$,
in which
\item[-] $s, s'$ are constructed from $t_1,\ldots,t_n$ by applying 
function symbols from $\mathit{Fun}_P$.
\end{enumerate}
$\mathit{R_p}$ is a \emph{valid} interargument relation for $\mathit{p}$ in $P$ w.r.t.\
$(\succsim,\succ)$ if and only if for every $p(t_1,\ldots,t_n) \in \mathit{Atom}_P$: $P
\models p(t_1,\ldots,t_n)$ implies $p(t_1,\ldots,t_n) \in R_p$.
\end{definition}

\begin{example}[interargument relation]\label{exam:interagr}
Let $P$ be the 
standard $\mathit{append}$ program that computes list concatenation. Then
there are a number of valid interargument
relations. Consider the reduction
pair $(\succsim,\succ)$ corresponding to the list-length norm ${\parallel}.{\parallel}_\ell$,
i.e., $t_1 \succsim t_2$ if and only if ${\parallel}t_1{\parallel}_\ell \ge
{\parallel}t_2{\parallel}_\ell$ and $t_1 \succ t_2$ if and only if
${\parallel}t_1{\parallel}_\ell > {\parallel}t_2{\parallel}_\ell$.
For instance, valid 
interargument relations for $append$ w.r.t. $(\succsim,\succ)$ are $R_{append}$
$=$ $\{append(t_1,t_2,t_3) \mid t_1,t_2,t_3 \in \mathit{Term}_P \wedge
\varphi_{append}(t_1,t_2,t_3) \}$, where $\varphi_{append}(t_1,t_2,t_3)$ could be:
\begin{enumerate}
\item[-] $t_3 \succsim t_2 \wedge t_3 \succsim t_1$,
\item[-] $t_3 \succsim t_2$,
\item[-] $[t_1,t_2|t_3] \succ [t_2|t_3]$, or
\item[-] $\mathit{true}$
\end{enumerate}
Of course, usually only the first two interargument relations
are useful for termination analysis.
{\hfill{$\mathit{\square}$}}
\end{example}

Finally, we need the notion of \emph{rigidity}, in order to deal with
bindings that are due to 
unification in LD-derivations. These bindings would have to be back-propagated
to the variables in the initial goal. We reformulate rigidity for arbitrary reduction
pairs.

\begin{definition}[rigidity - adapted from \cite{DeSchreyeSerebrenik01}] \label{def:rigidity}
A term or atom $A \in \mathit{Term}_P \cup \mathit{Atom}_P$ is called \emph{rigid}
w.r.t.\ a reduction pair $(\succsim,\succ)$ if $A \approx A \sigma$ holds for any
substitution $\sigma$. A set of terms (or atoms) $S$ is called \emph{rigid}
w.r.t.\ $(\succsim,\succ)$ if all its elements are rigid w.r.t.\ $(\succsim,\succ)$.
\end{definition}

\begin{example}[rigidity]\label{exam:rigidity}
The list $[X|t]$ ($X$ is a variable, $t$ is a ground term) is rigid w.r.t.\ the reduction
pair $(\succsim,\succ)$ corresponding to the list-length norm.
For any substitution
$\sigma$, we have ${\parallel}[X|t]{\sigma}{\parallel}_\ell =
1+{\parallel}t{\parallel}_\ell = {\parallel}[X|t]{\parallel}_\ell$. Therefore,
$[X|t]\sigma \approx [X|t]$ w.r.t. $(\succsim,\succ)$.

However, the list $[X|t]$ is not rigid w.r.t.\ the reduction pair $(\succsim',\succ')$
corresponding to the term-size norm ${\parallel}.{\parallel}_\tau$, i.e., $t_1 \succsim' t_2$
if and only if ${\parallel}t_1{\parallel}_\tau \ge
{\parallel}t_2{\parallel}_\tau$ and $t_1
\succ' t_2$ if and only if ${\parallel}t_1{\parallel}_\tau
> {\parallel}t_2{\parallel}_\tau$. 
{\hfill{$\mathit{\square}$}}
\end{example}

The following definition introduces the desired termination criterion, i.e., it
recalls the definition of \emph{rigid order-acceptability} w.r.t.\ a set of atoms.

\begin{definition}[rigid order-acceptability
\cite{DeSchreyeSerebrenik01}]\label{def:rigidacceptability} 
Let $S$ be a set of atomic queries. A  program $P$ is
\emph{rigid order-acceptable} w.r.t.\ $S$ if there exists a reduction pair $(\succsim,\succ)$
on $\mathit{Atom}_P$ 
where $\mathit{Call}(P,S)$ is rigid w.r.t.\ $(\succsim,\succ)$ and
where for each predicate $p$ in $P$, there is a valid  interargument
relation $R_{p}$ in $P$
w.r.t.\ $(\succsim,\succ)$ such that
%
\begin{enumerate}
\item[-] for any clause $A \incase B_1,B_2,\ldots,B_n$ in $P$,
\item[-] for any atom $B_i \in \{B_1\ldots,B_n\}$ such that $\mathit{rel}(B_i) \backsimeq \mathit{rel}(A)$,
\item[-] for any substitution $\theta$ such that the atoms
$B_1\theta, \ldots, B_{i-1}\theta$ are elements of their associated
interargument relations $R_{\mathit{rel}(B_1)}, \ldots,  R_{\mathit{rel}(B_{i-1})}$: 
                    \center{$A\theta \succ B_i\theta$}.
\end{enumerate}
\end{definition}

Theorem \ref{thm:rigidacceptability} states that 
rigid order-acceptability is a sufficient condition for termination. 
We refer to \cite{Serebrenik03}, Theorems 3.32 and 3.54, for the 
proof of Theorem \ref{thm:rigidacceptability}.

\begin{theorem}[termination criterion by rigid order-acceptability]
\label{thm:rigidacceptability} If $P$ is rigid
order-acceptable w.r.t.\ $S$, then $P$ terminates for any query in $S$.
\end{theorem}

Rigid order-acceptability is sufficient for
termination, but is not necessary for it (see
\cite{DeSchreyeSerebrenik01}). With Definition \ref{def:rigidacceptability}
and Theorem \ref{thm:rigidacceptability}, proving termination of a program
requires verifying the rigidity of the call set, verifying the validity of interargument
relations for predicates,  and verifying the decrease conditions for the (mutually)
recursive clauses. 

We will not discuss here the decidability or undecidability results
related to various problems concerning: (i) the rigidity of the call set
and (ii) the validity of interargument relations. The interested reader may refer
to the relevant literature.

In the remainder of this paper we provide some answers to the question in the setting of
a given set $S$, an inferred order based on polynomial interpretations, abstractions
of $S$ based on types, type inference to approximate the call set, and
interargument relations
based on inequalities between polynomials. 

\section{Polynomial Interpretation of a Logic Program}\label{sec:interpretation}
%
%

The approach presented in the previous section can be considered a theoretical
framework for termination analysis of LPs based on general
orders on terms and atoms. In this section, we specialize it to orders
based on polynomial interpretations. 

We first introduce polynomial interpretations in Section \ref{poly_interpretation}. Then in
Section \ref{Termination poly_interpretation} we reformulate the termination
conditions for LPs from Section \ref{subsec:conditions} for polynomial interpretations.

\subsection{Polynomial Interpretations}
\label{poly_interpretation}

In this paper, we only consider polynomials with natural numbers as
coefficients (so-called ``natural coefficients'').
Because natural numbers will occur many times in this paper, we will 
simply refer to them as ``numbers''.

We say that a variable $X$ \emph{occurs} in a polynomial $p$ if the polynomial
contains a monomial with a coefficient different from 0 and $X$ occurs in
this monomial.
If $X_1,\ldots,X_n$ are all the variables occurring in a polynomial $p$, we often
denote $p$ as $p(X_1,\ldots,X_n)$. For every polynomial $p$,
there is an associated polynomial function $F_p = \lambda X_1,\ldots,X_n.$ $p(X_1,\ldots,X_n)$.
For numbers or polynomials $x_1,\ldots,x_n$, we often write
``$p(x_1,\ldots,x_n)$''
instead of ``$F_p(x_1,\ldots,x_n)$''.
Given $p(X_1,\ldots,X_n)$ and $m \geq 1$ we also have 
an associated polynomial function $F_{p,m} = \lambda
X_1,\ldots,X_n,Y_1,\ldots,Y_m.$ $\ p(X_1,\ldots,X_n)$. For such 
an associated function on an extended domain, we often write 
``$p(x_1,\ldots,x_n,y_1,\ldots,y_m)$'' to denote 
``$F_{p,m}(x_1,\ldots,x_n,y_1,\ldots,y_m)$''.

\begin{definition}[orders on polynomials]
\label{ordersonnaturals}
Let $p$ and $q$ be two polynomials.
Let $X_1,\ldots,X_n$ be all variables occurring
in $p$ or $q$. The quasi-order
$\succsim_{\mathbb{N}}$  is defined as
$p \succsim_{\mathbb{N}} q$ if and only if $p(x_1,\ldots,x_n) \ge q(x_1,\ldots,x_n)$ for
all $x_1,\ldots,x_n \in \mathbb{N}$.
The strict order $\succ_{\mathbb{N}}$
is defined as $p \succ_{\mathbb{N}} q$ if and only
if  $p(x_1,\ldots,x_n) > q(x_1,\ldots,x_n)$ for
all $x_1,\ldots,x_n \in \mathbb{N}$.
\end{definition}




Observe that $(\succsim_{\mathbb{N}}, \succ_{\mathbb{N}})$ is a reduction pair.
In other words, 
$\succ_{\mathbb{N}}$ is well-founded and transitive,
$\succsim_{\mathbb{N}}$ is reflexive and transitive, and $\succsim_{\mathbb{N}}$ and 
$\succ_{\mathbb{N}}$ are compatible.
Let $\mathit{\Sigma}$ we denote the set of all polynomials with natural
coefficients.  Note that all these polynomials $p$ are \emph{weakly
monotonic}, i.e., $x_i \ge y_i$ for all $1 \leq i \leq n$ implies
$p(x_1,\ldots,x_n) \ge p(y_1,\ldots,y_n)$.

A \emph{polynomial interpretation} maps
each function and each predicate symbol of the program to a polynomial.

\begin{definition}[polynomial interpretation]
\label{def:interpretation}
A \emph{polynomial interpretation} $\mathit{I}$ for a logic program $P$
%
maps each 
symbol $f$ of arity $n$ in $\mathit{Fun}_P \cup \mathit{Pred}_P$ to a
polynomial $p_f(X_1,\ldots,X_n)$.
\end{definition}

Every polynomial interpretation induces a norm and a level mapping.
Although it is standard in logic programming to distinguish between norms and level
mappings, to simplify the formalization, here
we will only introduce a level mapping and define it
on both terms and atoms.

\begin{definition}[polynomial level mapping]
\label{def:polynomiallevelmapping}
The \emph{level mapping} associated with a polynomial interpretation $I$, 
is a mapping ${\mid}.{\mid}_{I}: \mathit{Term}_P \cup \mathit{Atom}_P \rightarrow \Sigma$,
which is defined recursively as: 
\begin{enumerate}
    \item[-] ${\mid}X{\mid}_{I} = X$ if $X$ is a variable,
    \item[-] ${\mid}f(t_1,\ldots,t_n){\mid}_{I} =
p_f({\mid}t_{1}{\mid}_{I},\ldots,{\mid}t_{n}{\mid}_{I})$,
where $p_{f} = I(f)$. 
\end{enumerate}
\end{definition}

\noindent
Every polynomial interpretation induces corresponding orders.

\begin{definition}[reduction pair corresponding to polynomial interpretation]\label{def:order-on-atom}
Let $I$ be a polynomial
interpretation. We define the relations $\succsim_I$ and $\succ_I$ on
$\mathit{Term}_P \cup \mathit{Atom}_P$ as follows:
\begin{enumerate}
\item[-] $s \succsim_I t$ if and only if
${\mid}s{\mid}_I \succsim_{\mathbb{N}} {\mid}t{\mid}_I$ for any
$s,t \in \mathit{Term}_P \cup \mathit{Atom}_P$
\item[-] $s \succ_I t$ if and only if
${\mid}s{\mid}_I \succ_{\mathbb{N}} {\mid}t{\mid}_I$ for any
$s,t \in \mathit{Term}_P \cup \mathit{Atom}_P$
\end{enumerate}
\end{definition}

Again, observe that the orders induced by a polynomial interpretation form a reduction pair.



\begin{example}[polynomial interpretation for ``\textit{der}'']\label{exam:dist}
Let $I$ be a polynomial interpretation with 
\[\begin{array}{llllclcll}
I(+) &=& I(*) &=&
p_{+}(X_1,X_2) &=&
p_{*}(X_1,X_2)&=& X_1 + X_2 + 2\\
I(u) &=& I(1)& =& p_u &=& p_1 &=& 1\\
\multicolumn{3}{c}{I(\mathit{der})} &=&
\multicolumn{3}{c}{p_{\mathit{der}}(X)} &=& X^2 + 2 X + 2\\
\multicolumn{3}{c}{I(d)}&=&\multicolumn{3}{c}{p_d(X_1,X_2)} &=& X_1
\end{array}\]
Then $d(\mathit{der}(X+Y), DX + DY) \succ_I d(\mathit{der}(X), DX)$, since
${\mid}d(\mathit{der}(X+Y), DX + DY){\mid}_I =  (X+Y+2)^2 + 2 (X + Y + 2) + 2
\succ_{\mathbb{N}} {\mid}d(\mathit{der}(X), DX){\mid}_I =  X^2 + 2 X + 2$.
\end{example}

\subsection{Termination of Logic Programs by Polynomial Interpretations}
\label{Termination poly_interpretation}

We now re-state Definition \ref{def:rigidacceptability} and Theorem
\ref{thm:rigidacceptability} for the special case of 
polynomial interpretations.
So instead of interargument relations for arbitrary orders as in Definition
\ref{def:interargument}, 
we now use
interargument relations \emph{w.r.t.\ polynomial interpretations}.

\begin{definition}[interargument relation w.r.t.\ a polynomial interpretation]
\label{def:polynomialinterargumentrelation}
Let $P$ be a program, $p$ be a
predicate in $P$, and $I$ be a polynomial interpretation. 
$R_{p}$ is an \emph{interargument relation}  for $p$ in $P$ w.r.t.\
$I$
iff $R_p$ is an interargument relation  for $p$ in $P$ w.r.t.\
$(\succsim_I,\succ_I)$.
\end{definition}


Instead of rigidity w.r.t.\ general orders as in Definition
\ref{def:rigidity}, we define \emph{rigidity w.r.t.\ polynomial interpretations}.

%

\begin{definition}[rigidity w.r.t.\ a polynomial interpretation]
\label{def:rigidity-preinterpretation}
A term or atom $A \in \mathit{Term}_P \cup \mathit{Atom}_P$ is called
\emph{rigid w.r.t.\ a polynomial interpretation}
$I$ iff
$A$ is rigid w.r.t.\ $(\succsim_I,\succ_I)$, i.e., 
iff 
$A\; {\approx}_I \, A \sigma$ holds for any
substitution $\sigma$. A set of terms (or atoms) $S$ is called \emph{rigid}
w.r.t.\ $I$ if all its elements are rigid w.r.t.\ $I$.
\end{definition}
%

For polynomial interpretations, rigidity can also be characterized in an
alternative way using \emph{relevant variables}.

\begin{definition}[relevant variables]\label{def:relevantvar}
Let $I$ be a polynomial interpretation and $A$ be a term or
atom. A variable $X$ in
$A$ is called \emph{relevant} w.r.t.\ $I$ if there exists a substitution
$\{X \rightarrow t\}$ of a term $t$ for $X$,
such that
$A\{X \rightarrow t\}
\not\approx_I A$. 
\end{definition}

\begin{example}[relevant variables]
Let $A=[X|Y]$ and $\mathit{I}$ be the interpretation corresponding to the
list-length norm ${\parallel}.{\parallel}_\ell$,
i.e., ${\mid}[H|T]{\mid}_I= 1+{\mid}T{\mid}_I$. Then
the only relevant variable of $A$ is $Y$.{\hfill{$\square$}} 
\end{example}

\begin{proposition}[alternative characterization of rigidity]\label{prop:relevantvar}
Let $I$ be a polynomial interpretation and $A$ be a term or atom. 
Then $A$ is rigid w.r.t.\
$I$ iff $A$ has no relevant variables w.r.t.\ $I$.
\end{proposition}
\begin{proof}\label{proof:relevantvar}
Obvious from Definitions \ref{def:rigidity-preinterpretation} and  \ref{def:relevantvar}. 
\end{proof}

Using the notions of interargument relations and 
rigidity w.r.t.\ a
polynomial 
interpretation, we obtain the following specialization of Theorem
\ref{thm:rigidacceptability}: 

\begin{corollary}[termination criterion with polynomial rigid order-ac\-ceptability]
\label{prop:polynomialacceptability}
    Let $S$ be a set of atomic queries and $P$ be a program. Let $I$ be a
polynomial interpretation,
where $\mathit{Call}(P,S)$ is rigid w.r.t. $I$
and where for each predicate $p$ in $P$,
there is a valid interargument relation 
$R_{p}$ in $P$ w.r.t.\
$I$ such that 
\begin{enumerate}
    \item[-] for any clause $A \incase B_1,B_2,\ldots,B_n$ in $P$,
    \item[-] for any atom $B_i \in \{B_1\ldots,B_n\}$  such that $\mathit{rel}(B_i) \backsimeq \mathit{rel}(A)$,
    \item[-] for any substitution $\theta$ such that the atoms
	  $B_1\theta, \ldots, B_{i-1}\theta$ are elements of their associated
interargument relations $R_{\mathit{rel}(B_1)}, \ldots,
R_{\mathit{rel}(B_{i-1})}$: \\
   \item[] \begin{center}$A\theta \succ_I B_i\theta$.
 \end{center}
\end{enumerate}
Then $P$ terminates for any query in  $S$.
\end{corollary}
\begin{proof}
The corollary immediately follows from 
Theorem \ref{thm:rigidacceptability}.
\end{proof}

Corollary \ref{prop:polynomialacceptability} can be applied to verify termination of a
logic program w.r.t.\ a set of queries. More precisely, we have to check that all conditions
in the following termination proof procedure are satisfied by some polynomial interpretation 
$I$. In Section \ref{sec:automation} we will discuss how to find such an interpretation
automatically.

\begin{procedure}[a procedure for automatic termination analysis]
The termination proof procedure derived from Corollary \ref{prop:polynomialacceptability}
contains the following three steps:
	\label{proof_procedure}
	\begin{enumerate}
		\item[] \textbf{\underline{Step 1}:} The call set $\mathit{Call}(P,S)$ must
be rigid w.r.t.\ $I$.  In other words, no query $A$ in the call set may have a relevant
variable w.r.t.\ $I$. 
		%
		%
		\item[] \textbf{\underline{Step 2}:} For a clause that has 
body-atoms between the head and a (mutually) recursive body-atom,  valid interargument
relations of those atoms w.r.t.\ $I$ need to be inferred. 
		%
		\item[] \textbf{\underline{Step 3}:} For every clause, the polynomial level
mapping of the head w.r.t.\ $I$ should be larger than that of any (mutually) recursive
body-atom, given that interargument relations for intermediate body-atoms hold.  
	\end{enumerate}
\end{procedure}

For Step 2, we can follow the standard approach for LPs to verify that a
relation $R$
holds for all elements
of the Herbrand model (see e.g. \cite{Lloyd87}). To this end, 
one has to verify $T_P(R) \subseteq R$, where
$T_P$ is the immediate consequence operator corresponding to the program $P$.
Thus, we verify the validity of interargument
relations by first checking whether they are correct for the facts in the program. Then for every
clause, if the interargument relations hold for all body-atoms, the
interargument
relation for the head
should also hold.

\begin{example}[applying Corollary \ref{prop:polynomialacceptability} to the
``\textit{der}''-program]\label{rigid-order:der}
Consider again the ``\textit{der}''-program from Example \ref{exam:der}
and the  set of queries $S=\{d(t_1,t_2) \mid t_1$ is a ground
term and
$t_2$ is an arbitrary term\}. Note that here, $\mathit{Call}(P,S) = S$.
Let $I$ be the polynomial interpretation from Example \ref{exam:dist}. 
%
%
%
Then no $A \in \mathit{Call}(P,S)$ has a relevant variable w.r.t.\
$I$.
This
means that $\mathit{Call}(P,S)$ is rigid w.r.t.\ $I$.

Let $R_{d} = \{d(t_1,t_2) \mid t_1, t_2 \in Term_P, 
t_1 \succ_I t_2\}$ be an
interargument relation for the predicate $d$.
Checking the validity of $R_{d}$ is equivalent to
verifying the correctness of the following conditions for any substitution
$\theta$:
\begin{center}
    $\mathit{der}(u)\theta \succ_I (1)\theta$\\\vspace{3mm} 
    $\mathit{der}(X)\theta \succ_I \mathit{DX}\theta$ and 
$\mathit{der}(Y)\theta \succ_I \mathit{DY}\theta$ implies \\  
    $\mathit{der}(X+Y)\theta \succ_I (\mathit{DX}+\mathit{DY})\theta$\\\vspace{3mm} 

    $\mathit{der}(X)\theta \succ_I \mathit{DX}\theta$ and 
$\mathit{der}(Y)\theta \succ_I \mathit{DY}\theta$ implies \\ 
$\mathit{der}(X*Y)\theta \succ_I (X*\mathit{DY}+Y*\mathit{DX})\theta$\\\vspace{3mm}
    $\mathit{der}(X)\theta \succ_I \mathit{DX}\theta$ and 
$\mathit{der}(\mathit{DX})\theta \succ_I \mathit{DDX}\theta$ implies\\
    $\mathit{der}(\mathit{der}(X))\theta \succ_I \mathit{DDX}\theta$.\\\vspace{3mm}
\end{center}

To prove termination, we also need the following decrease conditions for any substitution
$\theta$:
\begin{center}
    $d(\mathit{der}(X+Y),\mathit{DX}+\mathit{DY})\theta  \succ_I
d(\mathit{der}(X),\mathit{DX})\theta$\\\vspace{2mm}
    $d(\mathit{der}(X),\mathit{DX})\theta$ satisfies $R_{d}$ implies \\
$d(\mathit{der}(X+Y),\mathit{DX}+\mathit{DY})\theta \succ_I d(\mathit{der}(Y),\mathit{DY})\theta$\\\vspace{2mm}
    $d(\mathit{der}(X*Y),X*\mathit{DY}+Y*\mathit{DX})\theta \succ_I d(\mathit{der}(X),\mathit{DX})\theta$\\\vspace{2mm}
    $d(\mathit{der}(X),\mathit{DX})\theta$ satisfies $R_{d}$ implies \\
$d(\mathit{der}(X*Y),X*\mathit{DY}+Y*\mathit{DX})\theta \succ_I 
d(\mathit{der}(Y),\mathit{DY})\theta$\\\vspace{2mm}
    $d(\mathit{der}(\mathit{der}(X)),\mathit{DDX})\theta \succ_I d(\mathit{der}(X),\mathit{DX})\theta$\\\vspace{2mm}
    $d(\mathit{der}(X),\mathit{DX})\theta$ satisfies $R_{d}$ implies \\
$d(\mathit{der}(\mathit{der}(X)),\mathit{DDX})\theta \succ_D d(\mathit{der}(\mathit{DX}),\mathit{DDX})\theta$
\end{center}

The conditions above 
are equivalent to the following inequalities on the
variables $X, Y, \mathit{DX}, \mathit{DY}, \mathit{DDX}$. For the conditions on the valid interargument
relation, we obtain:

\vspace*{-.4cm}

{\small
\[\begin{array}{rcl}
   \hspace*{8cm} 5 &>& 1 \vspace*{.2cm}\\
\multicolumn{3}{l}{\forall X,Y,\mathit{DX},\mathit{DY} \in \mathbb{N}: \; X^2 +2 X + 2 > \mathit{DX}
\wedge Y^2 + 2 Y + 2
 > \mathit{DY} \Rightarrow}\\
 (X+Y+2)^2 + 2 (X + Y + 2) + 2 &>&
\mathit{DX}+\mathit{DY} + 2 \vspace*{.2cm} \\
\multicolumn{3}{l}{\forall X,Y,\mathit{DX},\mathit{DY} \in \mathbb{N}: \;
X^2 +2 X + 2 > \mathit{DX}
\wedge Y^2 + 2 Y + 2
 > \mathit{DY} \Rightarrow}\\
 (X+Y+2)^2 + 2 (X + Y + 2) + 2 &>&
x + \mathit{DY}+ Y + \mathit{DX} + 3 \vspace*{.2cm} \\
 \multicolumn{3}{l}{\forall X,\mathit{DX},\mathit{DDX} \in \mathbb{N}: \;
X^2 +2 X + 2 > \mathit{DX}
\wedge
\mathit{DX}^2 +2 \mathit{DX} + 2 > \mathit{DDX}  \Rightarrow}\\
(X^2 + 2 X + 2)^2 + 2 (X^2 + 2 X + 2) + 2
&>& DDX
\end{array}\]}

\vspace*{-.2cm}

\noindent
And for the decrease conditions we obtain:

\vspace*{-.3cm}

{\scriptsize
\[\begin{array}{rcl}
 \forall X,Y \in \mathbb{N}: \, (X+Y+2)^2 + 2 (X + Y + 2 ) + 2 &>& X^2 + 2 X + 2 \\
 \forall X,Y,\mathit{DX} \in \mathbb{N}: \,
 X^2 + 2 X + 2 > \mathit{DX} \;
\Rightarrow \;
(X+Y+ 2)^2 + 2 (X + Y + 2) + 2&>& Y^2 + 2 Y + 2\\
 \forall X,Y \in \mathbb{N}:  \, (X+Y+2)^2 + 2 (X + Y + 2 ) + 2 &>& X^2 + 2 X + 2\\
\forall X,Y,\mathit{DX} \in  \mathbb{N}: \,
X^2 + 2 X + 2 > \mathit{DX} \;
\Rightarrow \;
(X+Y+ 2)^2 + 2 (X + Y + 2) + 2&>& Y^2 + 2 Y + 2\\
 \forall X \in\mathbb{N}: \,
(X^2 + 2 X + 2)^2 + 2 (X^2 + 2 X + 2) + 2 
&>& X^2 + 2 X + 2\\
 \forall X,\mathit{DX} \in \mathbb{N}: \,
X^2 + 2 X + 2 > \mathit{DX} \;\Rightarrow \;
(X^2 + 2 X + 2)^2 + 2 (X^2 + 2 X + 2) + 2 &>& \mathit{DX}^2 + 2 \mathit{DX} + 2
\end{array}\]}

\vspace*{-.2cm}

The above inequalities are easily
verified for all instantiations of the variables by numbers. 
Hence, the program terminates w.r.t.\ the set of queries $S$. {\hfill{$\square$}}
\end{example}

\section{Automating the Termination Proof}
\label{sec:automation}

A key question is how to automate the search for a polynomial interpretation
and for interargument relations. In other words, to prove
termination of a logic program, one has to synthesize the coefficients of the polynomials
associated with the function and predicate symbols as well as the formulas
$\varphi_p(t_1,\ldots,t_n)$ defining the 
interargument relations.
In the philosophy of the constraint-based approach in
\cite{Decorteetal98}, we do not choose a particular polynomial interpretation
and particular interargument relations. Instead, we introduce a general symbolic form
for the polynomials associated with the function and predicate symbols and for
the interargument relations. As 
an example,  assume that polynomials of degree 2 are selected for the
interpretation. Then instead of assigning the polynomial $p_q(X_1,X_2) =
 X_1^2 + 2 X_1 X_2$ to a predicate symbol $q$ of arity 2, we would, for example, assign
the \emph{symbolic} polynomial $p_q(X_1,X_2) = q_{00} + q_{10} X_1 + q_{01}
X_2 
+ q_{11} X_1 X_2 + q_1 X_1^2 + q_2 X_2^2$,
where the $q_i$ and $q_{ij}$ 
are unknown coefficients ranging
over $\mathbb{N}$. So our approach for termination analysis works as follows:
\begin{itemize}
\item   introduce symbolic versions of the 
polynomials associated with function and predicate symbols, 
\item   express all conditions resulting from Corollary
\ref{prop:polynomialacceptability} as constraints on the coefficients
(e.g. $q_{00}, q_{10}, q_{01}, \ldots$), 
\item   solve the resulting system of constraints to obtain values for the coefficients.
\end{itemize}
Each solution for this constraint system gives rise to a concrete polynomial
interpretation and to concrete valid interargument
relations such that all conditions of 
Corollary
\ref{prop:polynomialacceptability} are satisfied. 
Therefore, each solution gives a termination proof. 

In order to assign symbolic polynomials to the function and predicate symbols,
we make the decision of assigning linear polynomials to predicate symbols
and linear or simple-mixed polynomials to function symbols.
These classes of polynomials are defined as follows:
\begin{enumerate}
\item[-] \emph{The linear class:} each monomial of a polynomial in this class
contains at most one variable of at most degree 1:\\
$p(X_1,\ldots,X_n)=p_0+\sum_{k=1}^{n} p_k X_k$  \vspace*{.2cm}
\item[-] \emph{The simple-mixed class:} each monomial of a polynomial in this
class contains either a single variable of at most degree 2 or several
variables of at most degree 1:\\ 
$p(X_1,\ldots,X_n)=\sum_{j_k \in \{0,1\}} p_{j_1{\ldots}j_n} X^{j_1}_1 \ldots X^{j_n}_n
+ \sum_{k=1}^{n} p_k X^2_k$
\end{enumerate}
The above classes of polynomials have proved to be particularly useful for
automated termination proofs of TRSs. 
For more details on these classes of polynomials we refer to
\cite{contejean05jar,Steinbach92}.
In our work, these choices resulted from extensive experiments
with different kinds of polynomials, where our goal was to optimize both the
efficiency and the power of the termination analyzer.

In Section \ref{Reformulating 
the Termination Conditions Symbolically}, we first reformulate the conditions
of our termination criterion in  Corollary
\ref{prop:polynomialacceptability}, using the above symbolic forms of
polynomials. Then in Section \ref{rewriting}, we transform these symbolic conditions
into constraints on the unknown coefficients of the symbolic polynomials. Afterwards, 
in Section \ref{Solving Diophantine Constraints} we show how these resulting
Diophantine constraints 
can be solved automatically. Finally, we conclude with a comparison of our
contributions with related work from term rewriting in Section \ref{Discussion}.

\subsection{Reformulating the Termination Conditions}\label{Reformulating 
the Termination Conditions Symbolically}

In this subsection, we reformulate all termination conditions of Corollary
\ref{prop:polynomialacceptability}, i.e., of Procedure \ref{proof_procedure}. 
These include the rigidity property (Step 1),
the valid interargument relations (Step 2), and the
decrease conditions (Step 3). The reformulation results in
symbolic constraints, based on the
symbolic forms of the polynomial interpretations.

\subsubsection{Rigidity Conditions (Procedure \ref{proof_procedure}, Step 1)}\label{Rigidity Conditions}\hspace*{\fill}

\vspace*{.2cm}


\noindent
There are several ways to approximate $\mathit{Call}(P,S)$ (e.g.,
\cite{Bruynoogheetal05,GallagherHB05,HeatonACK00,Janssensetal92}). 
%
%
%
In this paper, we apply the approximation technique
of \cite{GallagherHB05,Janssensetal92}. More precisely, we first
%
%
specify the set of queries as a set of rigid
type graphs. 
Then the technique in \cite{GallagherHB05,Janssensetal92} is used to 
compute a new, finite set of rigid type graphs which 
approximate
$\mathit{Call}(P,S)$. Each of these new rigid type graphs represents
 a so-called
call pattern.
For further details, we
refer to \cite{GallagherHB05,Janssensetal92}.  

In the following, we 
recapitulate the notion of rigid type graphs
and show how rigidity conditions are derived from the set of call patterns. First,
we recall and extend some basic definitions from 
\cite{Janssensetal92}, which are based on linear norms and
level-mappings, to the case of general polynomial
interpretations.  Example \ref{der:symbolrigidcond} will illustrate these definitions.

%

\begin{definition}[rigid type graph \cite{Janssensetal92}] 
\label{rigidtypegraph}
A rigid type graph $T$ is a 5-tuple, $(\mathit{Nodes}, \mathit{ForArcs},
\mathit{BackArcs}, \mathit{Label}, \mathit{ArgPos})$, where 
\begin{enumerate}
    \item[1.] $\mathit{Nodes}$ is a finite non-empty set of nodes.
    \item[2.] $\mathit{ForArcs} \subseteq \mathit{Nodes} \times
\mathit{Nodes}$ such that 
$(\mathit{Nodes},\mathit{ForArcs})$ is a tree.
    \item[3.] $\mathit{BackArcs} \subseteq \mathit{Nodes} \times
\mathit{Nodes}$
 such that for every
arc $(m,n) \in \mathit{BackArcs}$, node $n$ is an ancestor of node $m$ in the tree $(\mathit{Nodes},\mathit{ForArcs})$.
    \item[4.] $\mathit{Label}$ is a function $\mathit{Nodes} \rightarrow 
\mathit{Fun}_P 
\cup \mathit{Pred}_P \cup \{\mbox{\rm \textbf{MAX}},\mbox{\rm \textbf{OR}}\}$.
    \item[5.] If a node $n$ is labelled with $f \in \mathit{Fun}_P \cup
\mathit{Pred}_P$ and $f$ has arity $k$, then the node $n$ has exactly $k$ outgoing arcs (counting both $\mathit{ForArcs}$
and $\mathit{BackArcs}$). These arcs are labelled with the numbers
$1,\ldots,k$.  For every such arc $(n,m)$, $\mathit{ArgPos}(n,m)$ returns
the corresponding label from $\{1,\ldots,k\}$.
\end{enumerate}
\end{definition}

The intuition behind rigid type graphs is related to the tree representation 
of terms and atoms in LP. A rigid type graph generalizes the tree representation
of an atom by allowing:
\begin{itemize}
\item nodes labeled by \mbox{\rm \textbf{MAX}}, denoting any term,
\item nodes labeled by \mbox{\rm \textbf{OR}}, denoting the union of all denotations of the sub-graphs rooted at this node,
\item backarcs, denoting repeated traversals of a sub-graph.
\end{itemize}

For each rigid type graph representing a set of atoms $S$, each node $\mbox{\rm	\textbf{MAX}}$ in
the graph corresponds to a possible occurrence of a variable in the atoms of $S$. The
set $S$ is rigid w.r.t.\ the polynomial interpretation $I$ iff all these variables
are not relevant w.r.t.\ $I$. In the following, we formulate this
rigidity condition syntactically based on the rigid type graph.
\begin{definition}[critical path \protect\cite{Decorteetal98}]
\label{criticalpath}
    Let $T\!=\!(\mathit{Nodes}, \mathit{ForArcs}, \mathit{BackArcs},
\mathit{Label}, \mathit{ArgPos})$ be a rigid type graph. A 
critical path in $T$ is a path of arcs from the tree $\mathit{ForArcs}$ which goes from
the root node of the tree to a node 
labelled $\mbox{\rm \textbf{MAX}}$. 
\end{definition}
%

The following proposition is extended from \cite{Decorteetal93}, where in
\cite{Decorteetal93} each function or predicate symbol is associated with a
linear norm or level mapping. It provides a method to generate 
constraints for rigidity.

\begin{proposition}[checking rigidity by critical paths]
\label{rigiditysyntax}
    Let $P$ be a program and $T = (\mathit{Nodes}, \mathit{ForArcs}, \mathit{BackArcs}, \mathit{Label}, \mathit{ArgPos})$ be
a rigid type graph representing a set of atoms $S$. Let
$I$ be a polynomial
interpretation, where for any function or predicate symbol $f$ of arity $k$ we
have $I(f) = p_f(X_1,\ldots,X_k) =
\sum_{0 \le j_1,\ldots,j_k \le M_f} f_{{j_1}\ldots{j_k}}X_1^{j_1} \ldots X_k^{j_k}$. 
The set $S$ is rigid w.r.t.\ $I$ iff 
on every critical path of $T$ there exists an arc $(n, m)$
with $\mathit{Label}(n) = f$, $\mathit{arity}(f) = k$, and $\mathit{ArgPos}(n,m) = i$ such that
$\sum_{j_i>0} f_{{j_1}\ldots{j_k}} = 0$, where $k$ is the arity of $f$.
\end{proposition}
\begin{proof}
Since we only regard polynomials with non-negative
coefficients
$f_{{j_1}\ldots{j_k}}$, 
the condition $\sum_{j_i>0} f_{{j_1}\ldots{j_k}} = 0$ 
is equivalent to the requirement that $f_{{j_1}\ldots{j_k}} = 0$, whenever
$j_i>0$. This in turn is equivalent to the condition that $X_i$ is not 
involved in $p_f(X_1,\ldots,X_k)$. Hence, the condition in the above proposition
is equivalent to the requirement that
for any $\mbox{\rm	\textbf{MAX}}$
node, there is at least one function or predicate symbol $f$ on the critical path to this
$\mbox{\rm	\textbf{MAX}}$ node, for 
which the argument position corresponding to the path is not involved in
$p_f$. So equivalently, the atoms in the set $S$ have no relevant variables
w.r.t.\ $I$.
According to Proposition \ref{prop:relevantvar}, this is equivalent to rigidity 
w.r.t.\ $I$.
\end{proof}


The following corollary shows how to express the above rigidity check as a
constraint on the coefficients of the polynomial interpretation. To this end,
we express the existence condition of an appropriate arc $(n, m)$
by a suitable multiplication.

\begin{corollary}[symbolic condition for checking rigidity]
\label{corol-rigid2}
    Let $T$ be a rigid type graph
 representing a set of atoms $S$ and let $\mathit{CP}$ be a critical path of $T$. Let
$(n^1,m^1), \ldots, (n^e,m^e)$ be all arcs in
$\mathit{CP}$ such that for all $d \in \{1,\ldots,e\}$, 
$\mathit{Label}(n^d) = f^d$
is a function or predicate symbol of some arity $k^d$ and $\mathit{ArgPos}(n^d,\linebreak
m^d) = i^d$. 
If for any such $\mathit{CP}$ we have
 \begin{align} 
    \prod_{d=1}^{e} \quad (\sum_{j_{(i^d)} \, > \,0}
f^d_{{j_{1}}\ldots{j_{(k^d)}}}) = 0, \label{form:rigid} 
 \end{align}
then $S$ is rigid w.r.t.\ $I$.
\end{corollary}
%

        \begin{figure}[t]
                \centering
\begin{picture}(220,170)(10,100)
                \includegraphics[scale=0.40]{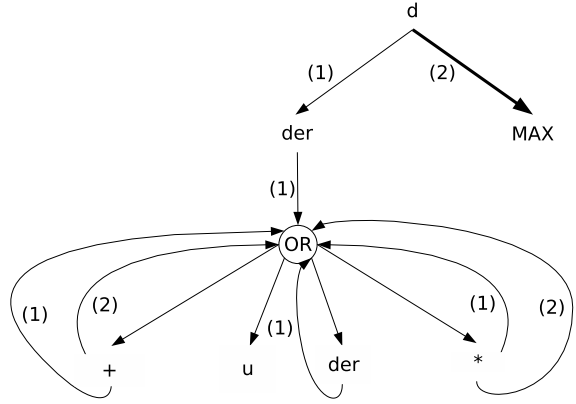}
\end{picture}
                \caption{Rigid type graph for Example \ref{der:symbolrigidcond}}
                \label{fig:example:rigidcondi-der}
        \end{figure}

\begin{example}[symbolic polynomial interpretation and 
rigidity constraints for the ``\textit{der}''-program]
\label{der:symbolrigidcond}
For Example \ref{exam:der},
we define a symbolic polynomial interpretation $I$ as 
follows.

%

\[\begin{array}{rcl}
I(+) &= &p_1 X_1^2+p_2 X_2^2+p_{11} X_1X_2+p_{10} X_1+p_{01} X_2+p_{00} \\ 
I(*) &= &m_1 X_1^2+m_2 X_2^2+m_{11} X_1X_2+m_{10} X_1+m_{01} X_2+m_{00} \\
I(\mathit{der}) &=&\mathit{der}_2 X^2+\mathit{der}_1 X+\mathit{der}_0 \\
I(u) &=& c_u \\
I(1) &=&c_{1} \\
I(d) &=&d_0 + d_1 X_1 + d_2 X_2
\end{array}\]

We will reformulate the termination conditions for this example in symbolic form. However
for reasons of space, we
will not give all polynomial constraints. Instead, in order to illustrate the main
ideas, in each sub-section we only present one constraint for the corresponding type of
conditions. 

Instead of checking termination of the ``$\mathit{der}$''-program w.r.t.\
the set of queries $S = \{d(t_1,t_2) \mid t_1$ is a ground term, $t_2$ is an
arbitrary term$\}$  
as in Example \ref{exam:der}, we now regard the set of queries $S_1 =
\{d(t_1,t_2) \mid t_1$ is of the form $der(t'_1)$,  
where $t'_1$ is a ground term constructed from the function symbols $u$, $+$, $*$,
$\mathit{der}$, and $t_2$ is an arbitrary term$ \}$. $S_1$ is represented by the type graph in
Figure \ref{fig:example:rigidcondi-der}. 

Obviously, termination of
the program w.r.t.\ $S_1$ also implies termination w.r.t. $S$. 
This can be proved easily by showing
		that for any query $Q \in S \setminus S_1$, the program trivially 
terminates by finite failure. 

In our example,
type inference \cite{Janssensetal92} computes the call set $\mathit{Call}(P,S_1)= S_1$,
i.e., the graph in Figure \ref{fig:example:rigidcondi-der} also represents
$\mathit{Call}(P,S_1)$. Its only critical path consists of just the arc from the root to the
node labelled \textbf{MAX}. Hence from 
the graph, the following rigidity
condition is generated according to Corollary \ref{corol-rigid2}: 
               \[d_2 = 0 \]
\hfill{$\square$}
\end{example}
\subsubsection{Valid Interargument Relations (Procedure \ref{proof_procedure}, Step 2)}
\label{Valid_Interargument_Relations}
\hspace*{\fill}

\vspace*{.2cm}

\noindent
Next we consider the other symbolic constraints, derived for valid interargument 
relations and decrease conditions. We will show that they all take the form:
\setcounter{polycons2ctr}{\value{equation}}
\begin{equation}
   \forall \overline{X} \in \mathbb{N}: \quad p_1 \ge q_1 \wedge \ldots \wedge
p_n \ge q_n \quad \Rightarrow \quad p_{n+1} \ge q_{n+1} \label{polycons2}
\end{equation}

\noindent
where $n \ge 0$ and $p_i, q_i$ are polynomials with natural
coefficients. Here, $\overline{X}$ is the tuple of all variables occurring in $p_1, \ldots,
p_{n+1}, q_1, \ldots, q_{n+1}$.

There are a number
of works on inferring valid interargument relations of predicates. In
\cite{Decorteetal98}, interargument relations are
formulated as inequalities between a linear
combination of the ``\textit{inputs}'' and a linear combination of the
``\textit{outputs}''. We will not define input and output arguments formally in this paper,
since we do not use them in our approach, but informally, inputs are the arguments of a
predicate symbol which are only called with ground terms and outputs are the remaining arguments.

We propose a new form of interargument relation, namely
\emph{polynomial}
interargument relations, which are of the following form: 
\begin{align}
    R_p = \{p(t_1,\ldots, t_n) \mid i_p({\mid}t_1{\mid}_I,\ldots,
{\mid}t_n{\mid}_I)   \succsim_\mathbb{N}  o_p({\mid}t_1{\mid}_I,\ldots,
{\mid}t_n{\mid}_I)\} \label{interarg} 
\end{align}
where $i_p$ and $o_p$ are polynomials with natural coefficients.

The form of interargument relations in \cite{Decorteetal98} can be
considered a special case of the form (\ref{interarg}) above, 
where $i_p({\mid}t_1{\mid}_I,\ldots,
{\mid}t_n{\mid}_I)$ is constructed from the
input arguments only and $o_p({\mid}t_1{\mid}_I,\ldots,
{\mid}t_n{\mid}_I)$ is only constructed from the
outputs.
%

Since the approach in \cite{Decorteetal98} only considers relations between
the input and output arguments of the predicates, it has
some limitations. In some cases, the desired relation does not compare
inputs with outputs, but the relation holds among the inputs only or among 
the outputs only. In particular, if all arguments of a predicate
are inputs (or outputs), then the approach in
\cite{Decorteetal98} fails to infer any useful relation among them. The following
example shows this point. It computes the natural division of the first and second
arguments of the predicate $\mathit{div}$ and returns the result in its
third argument.

\begin{example}[div]
\label{div}
\begin{align}
    & \mathit{div}(X, s(Y), 0) \incase \mathit{less}(X,s(Y)).   \nonumber          \\
    & \mathit{div}(X, s(Y), s(Z)) \incase
        \mathit{sub}(X, s(Y), R),
        \mathit{div}(R, s(Y), Z).            \label{div2}\\
  &\mathit{sub}(X, 0, X).               \nonumber\\
  & \mathit{sub}(s(X), s(Y), Z) \incase \mathit{sub}(X, Y, Z).    \nonumber\\
 &\mathit{less}(0,s(Y)).            \nonumber\\   
    & \mathit{less}(s(X),s(Y)) \incase \mathit{less}(X,Y).   \nonumber
\end{align}

We consider the set of queries   $S=\{\,
\mathit{div}(t_1,t_2,t_3) \mid t_1$ and $t_2$ are ground terms, and $t_3$ is an arbitrary 
term\}.
This program terminates for all these queries. If we look at Clause (\ref{div2}),
the decrease in size between the head and the recursive body-atom can be established if
we can infer a suitable valid interargument relation for $\mathit{sub}$. This relation 
should imply that within Clause (\ref{div2}), the 
first argument of $\mathit{sub}$ is greater than its third argument. However, if we apply the
approach in \cite{Decorteetal98}, inferring such an interargument relation for
$\mathit{sub}$ is impossible. Since  the
first two $\mathit{sub}$-arguments are used as input and the last one is output,
the approach can only infer interargument
relations where a linear combination of the sizes of the first
and second arguments is greater than or equal to the size of the third
argument. Then, we cannot conclude 
that for every successful answer substitution for the call $\mathit{sub}(X,s(Y),R)$ in Clause
(\ref{div2}), the first $\mathit{sub}$-argument $X$ is strictly greater than the 
third $\mathit{sub}$-argument $R$.

In contrast, if we use Form (\ref{interarg}), then it
is possible to infer the following valid interargument relation for $\mathit{sub}$: 
\[ R_{\mathit{sub}} = \{ \mathit{sub}(t_1,t_2,t_3) \mid
{\mid}t_1{\mid}_I \succsim_\mathbb{N}
{\mid}t_2{\mid}_I + {\mid}t_3{\mid}_I \}\] 
Note that in the right-hand side  ${\mid}t_2{\mid}_I + {\mid}t_3{\mid}_I$ of the 
above inequality, we have both an input argument $t_2$ and an output
argument $t_3$.
This valid polynomial interargument relation guarantees that for any
successful answer substitution for
the call $\mathit{sub}(X,s(Y),R)$ in Clause (\ref{div2}), we have
${\mid}X{\mid}_I \succ_\mathbb{N} {\mid}R{\mid}_I$ if
${\mid}s(Y){\mid}_I \succsim_\mathbb{N} 1$.
Our implementation in the system
\textsf{Polytool} is indeed able to infer this
interargument relation using the constraint solving technique explained below.
Therefore, \textsf{Polytool} can
prove termination of
``\textit{div}''. 
If
we used the form of interargument relations in \cite{Decorteetal98} instead,
\textsf{Polytool} would not be able to solve this problem.
\hfill{$\square$}.
\end{example}


Similar to the symbolic form of polynomial interpretations, we also use
a symbolic form of polynomial interargument relations. To this end,
we take
symbolic  polynomials $i_p$ and
$o_p$. For the inference of valid interargument relations, we then apply the technique
proposed in \cite{Decorteetal98}, cf.\ Procedure \ref{proof_procedure}, Step 2.
%
%
For any sequence of terms $t_1,\ldots, t_n$, let $\mathbf{R}_{p}(t_1,\ldots, t_n)$ abbreviate
the inequality  $i_p({\mid}t_1{\mid}_I,\ldots,
{\mid}t_n{\mid}_I)  \geq o_p({\mid}t_1{\mid}_I,\ldots,
{\mid}t_n{\mid}_I)$. 
The goal is to impose constraints on the polynomials $i_p$ and $o_p$ which
ensure that the corresponding interargument relation
$R_p = \{ p(t_1,\ldots,t_n) \mid   \forall \overline{X} \in \mathbb{N}: \,
\mathbf{R}_{p}(t_1,\ldots, t_n) \}$
is valid. To this end,
we generate for every clause of the program:
\[p(\overline{t}) \incase p_1(\overline{t_1}), \ldots,
p_n(\overline{t_n})\]
the constraint 
\[   \forall \overline{X} \in \mathbb{N}: \quad 
\mathbf{R}_{p_1}(\overline{t_1}) \wedge \ldots \wedge
\mathbf{R}_{p_n}(\overline{t_n}) \Rightarrow \mathbf{R}_{p}(\overline{t}). \]
It is clear that
this formula has Form (\ref{polycons2}).

\begin{example}[symbolic interargument relation for the 
``\textit{der}''-program]
\label{symbolic-der}
 We continue Example \ref{der:symbolrigidcond} and
use linear polynomials for 
$i_{\mathit{der}}$ and 
$o_{\mathit{der}}$, i.e., $i_{\mathit{der}}(X,Y) = i_0 + i_1 X+ i_2 Y$ and
$o_{\mathit{der}} = o_0+ o_1 X+ o_2 Y$. Hence, the 
the symbolic form of the polynomial interargument relation for the predicate  $d$ is
\begin{equation*}
    R_{d} = \{d(t_1,t_2) \mid i_0 + i_1{\mid}t_1{\mid}_I
+i_2{\mid}t_2{\mid}_I \succsim_\mathbb{N} o_0 +
o_1{\mid}t_1{\mid}_I +
o_2{\mid}t_2{\mid}_I\}. 
\end{equation*}
There are four clauses (\ref{der1}) -
(\ref{der4}) from which constraints for 
valid interargument relations are inferred. We only present the constraint
resulting from the last clause
(\ref{der4}):
\[d(\mathit{der}(\mathit{der}(X)),\mathit{DDX}) \incase 
d(\mathit{der}(X),\mathit{DX}), d(\mathit{der}(\mathit{DX}),\mathit{DDX})\]
Here, we obtain the constraint
\begin{eqnarray}
\nonumber && \forall X, \mathit{DX}, \mathit{DDX} \in
\mathbb{N}:\\
&&\mathbf{R}_{d}(\mathit{der}(X),\mathit{DX}) \wedge 
\mathbf{R}_{d}(\mathit{der}(\mathit{DX}),\mathit{DDX})  \Rightarrow 
\mathbf{R}_{d}(\mathit{der}(\mathit{der}(X)),\mathit{DDX}). \label{IAConstraint}
\end{eqnarray}
{\hfill{$\square$}}		
\end{example}
    %

\subsubsection{Decrease Conditions (Procedure \ref{proof_procedure}, Step
3)}\label{Decrease Conditions}\hspace*{\fill} 

\vspace*{.2cm}

\noindent
Finally, one has to require the decrease condition between the head and any
(mutually) recursive body-atom in any (mutually) recursive clause.
So for any clause 
\[p(\overline{t}) \incase p_1(\overline{t_1}), \ldots,
p_n(\overline{t_n})\]
of the program where 
$p \backsimeq p_i$ (i.e., where $p$ and $p_i$ are mutually recursive),
we require
\[   \forall \overline{X} \in \mathbb{N}: \quad 
\mathbf{R}_{p_1}(\overline{t_1}) \wedge \ldots \wedge
\mathbf{R}_{p_{i-1}}(\overline{t_{i-1}}) \Rightarrow
{\mid}p(\overline{t}){\mid}_I \geq {\mid}p_i(\overline{t_i}){\mid}_I +1.\]
Obviously, 
the formula is in Form 
(\ref{polycons2}).

\begin{example}[constraints for
the decrease conditions of  ``\textit{der}''] 
\label{der:symboldecreasecond}
	There are three recursive clauses  
(\ref{der2}) - (\ref{der4}) 
where decrease conditions can be
inferred. We present the decrease condition for the 
	recursive body-atom $d(\mathit{der}(\mathit{DX}),\mathit{DDX})$ of the
last clause (\ref{der4}):  
        	\begin{align}
        		&   \forall X, \mathit{DX}, \mathit{DDX} \in
						\mathbb{N}:
\nonumber       \\ 
        		&
i_0+i_1(\mathit{der}_2X^2+\mathit{der}_1X+\mathit{der}_0)+i_2\mathit{DX} \ge
\nonumber       \\ 
        		&
o_0+o_1(\mathit{der}_2X^2+\mathit{der}_1X+\mathit{der}_0)+ o_2\mathit{DX}
\nonumber       \\ 
        		&   \Rightarrow               \label{der:decreasecons}\\
        		&   d_0 +
			d_1(\mathit{der}_2(\mathit{der}_2X^2+\mathit{der}_1X+\mathit{der}_0)^2+ 
\nonumber \\  
			& \phantom{d_0 +
			d_1(}
			\mathit{der}_1(\mathit{der}_2X^2+\mathit{der}_1X
+\mathit{der}_0)+\mathit{der}_0)
			+d_2 \mathit{DDX}  \ge \nonumber\\
        		&   d_0 +
			d_1(\mathit{der}_2\mathit{DX}^2+\mathit{der}_1\mathit{DX}
+\mathit{der}_0)+
			d_2 \mathit{DDX} +1. \nonumber
        	\end{align}
{\hfill{$\square$}}
\end{example}

\subsection{From Symbolic Conditions to Constraints on Coefficients}
\label{rewriting}

Our goal is to find a polynomial interpretation such that all
constraints generated in the previous section are satisfied. To this
end, we transform all these constraints into Diophantine
constraints. In this transformation, we first eliminate
implications, cf.\ Section \ref{First Phase}. Afterwards, in Section \ref{poly-inter-generating},  
the universally quantified variables
(e.g., $X, DX, DDX, \ldots$) are removed
and the former unknown
coefficients (e.g., $\mathit{der}_0, \mathit{der}_1, \mathit{der}_2,  \ldots$)
become the new variables. If the resulting
Diophantine constraints can be solved, then the program under consideration is
terminating.

As we analyzed in Section \ref{Rigidity Conditions}, all generated rigidity
constraints have the Form (\ref{form:rigid}). Hence, these are already Diophantine
constraints which only contain unknown coefficients, but no universally
quantified variables. 

The other 
constraints, generated for the
valid interargument relations and the decrease conditions, have the following
form:
\setcounter{auxctr}{\value{equation}}
\setcounter{equation}{\value{polycons2ctr}}
 \begin{equation}
   \forall \overline{X} \in \mathbb{N}: \quad p_1 \ge q_1 \wedge \ldots \wedge
p_n \ge q_n  \quad \Rightarrow  \quad p_{n+1} \ge q_{n+1},
\end{equation}
where $n \ge 0$ and $p_i, q_i$ are polynomials with natural
coefficients.
\setcounter{equation}{\value{auxctr}}

In the following, we introduce a two-phase method to transform
all constraints of Form (\ref{polycons2}) into
Diophantine constraints on the unknown coefficients.

\subsubsection{First Phase: Removing Implications}\label{First Phase}\hspace*{\fill}

\vspace*{.2cm}

\noindent
The constraints of Form (\ref{polycons2}) are implications. In the first
phase, such constraints are transformed into inequalities without premises,
i.e., into constraints of the form
\setcounter{polycons1ctr}{\value{equation}}
 \begin{equation}
   \forall \overline{X} \in \mathbb{N}: \quad p \ge 0. \label{polycons1}
\end{equation}
However, here $p$ is a polynomial with integer (i.e., possibly negative)
coefficients. The transformation is \emph{sound}:
if the new constraints of Form
(\ref{polycons1}) are satisfied by some
substitution which instantiates the unknown coefficients with numbers, 
then this substitution also satisfies the
original constraints of Form (\ref{polycons2}).

The idea for the transformation is the following. Constraints
of the form (\ref{polycons2}) may have an arbitrary number $n$ of
premises $p_i \ge q_i$. We first transform them into
constraints with at most 
one premise. 
Obviously, $p_1 \ge q_1 \wedge \ldots \wedge
p_n \ge q_n$ implies $p_1 + \ldots + p_n \ge q_1 + \ldots q_n$. Thus, instead
of (\ref{polycons2}), it would be sufficient to demand
\[
   \forall \overline{X} \in \mathbb{N}: \quad p_1  + \ldots + p_n \ge q_1 + \ldots q_n
\quad \Rightarrow \quad p_{n+1} \ge q_{n+1}.
\]

So in order to combine the $n$ polynomials in the premise, we can use the
polynomial $\mathit{prem}(X_1,\ldots,X_n) = X_1 + \ldots + X_n$. Then instead
of (\ref{polycons2}), we may require
\[
   \forall \overline{X} \in \mathbb{N}: \quad \mathit{prem}(p_1,\ldots,p_n) \ge
\mathit{prem}(q_1,\ldots,q_n)
\quad \Rightarrow \quad p_{n+1} \ge q_{n+1}.
\]
A similar method
was also used  for termination
analysis of logic programs in \cite{Decorteetal98} and  for termination of term rewriting
in \cite[Section 7.2]{JAR07} to transform
disjunctions of polynomial inequalities into one single inequality.

For example, the constraint
\[ \forall X_1,X_2,X_3 \in \mathbb{N}: \quad X_1 \ge X_2 \wedge X_2 \ge  X_3
\quad \Rightarrow 
\quad X_1 \geq X_3\]
can now be transformed into
\[ \forall X_1,X_2,X_3 \in \mathbb{N}: \quad X_1 + X_2 \ge X_2 +  X_3
\quad \Rightarrow \quad X_1 \geq X_3\]
Since the latter constraint is valid, the former one is valid as well.

However, in order to make the approach more powerful, one could also use other
polynomials $\mathit{prem}$ in order to combine the $n$ inequalities in the
premise. The reason is that if $\mathit{prem}$ is restricted to be the
addition, then many valid constraints of the form (\ref{polycons2}) would be
transformed into invalid ones. For example, the valid constraint
\[ \forall X_1,X_2,X_3 \in \mathbb{N}: \quad X_1 \ge X_2^2 \wedge X_2 \ge X_3^2
\quad \Rightarrow \quad X_1  \geq X_3^4\]
would be transformed into the invalid constraint
\[ \forall X_1,X_2,X_3 \in \mathbb{N}: \quad X_1 + X_2 \ge X_2^2 +  X_3^2
\quad \Rightarrow \quad X_1 \geq X_3^4.\]
For instance, the constraint does not hold for
$X_1 = 4$, $X_2 = 0$, and $X_3 = 2$.

To make the transformation more general and more powerful,
we therefore permit
the use of \emph{arbitrary} polynomials $\mathit{prem}$ with natural
coefficients. In the above example, now the resulting constraint
\[ \forall X_1,X_2,X_3 \in \mathbb{N}: \quad \mathit{prem}(X_1,X_2) \ge \mathit{prem}(X_2^2,X_3^2)
\quad \Rightarrow \quad X_1 \geq X_3^4\]
would indeed be valid for a suitable choice of $\mathit{prem}$. For instance,
one could choose $\mathit{prem}$ to be the addition of the first argument with
the square of the second argument (i.e., $\mathit{prem}(X_1,X_2) = X_1 +
X^2_2$). 

By the introduction of the new polynomial $\mathit{prem}$, 
every constraint of the form (\ref{polycons2}) can now be transformed into an
implication with at most one premise. It remains to
transform such implications further into unconditional 
inequalities.
Obviously, instead of 
\begin{equation}
\label{version1}
\mathit{prem}(p_1,\ldots,p_n) \ge
\mathit{prem}(q_1,\ldots,q_n)
\; \Rightarrow \; p_{n+1} \ge q_{n+1},
\end{equation}
it is sufficient to demand 
\begin{equation}
\label{version2} p_{n+1} - q_{n+1} \ge
\mathit{prem}(p_1,\ldots,p_n) - \mathit{prem}(q_1,\ldots,q_n).
\end{equation}
This observation
was already used in the work of \cite{Decorteetal98} and also in termination
techniques for term rewriting  
to handle such conditional
polynomial inequalities
\cite{CADE98,CADE07}.

However, the approach can still be improved.
Recall that we used an arbitrary polynomial $\mathit{prem}$ to combine the
polynomials in the former premises. In a similar way, one could also apply an
arbitrary polynomial  $\mathit{conc}$ to the polynomials $p_{n+1}$ and
$q_{n+1}$ in
the former conclusion.
To see why this can be necessary, consider the valid constraint
\[ \forall X \in \mathbb{N}: \quad 2 X \geq 2 \quad \Rightarrow \quad X \geq 1.\]
With the transformation of (\ref{version1}) into 
(\ref{version2})
above, it would be transformed into the unconditional
constraint
\[ \forall X \in \mathbb{N}: \quad X -1 \; \geq \; 2 X - 2,\]
which is invalid. 
We have encountered several examples of this kind in our experiments, 
which motivates this further extension. In such examples, it 
would be better to apply a suitable
polynomial  $\mathit{conc}$ to the polynomials $X$ and $1$ in the former
conclusion. Then we would obtain
\[ \forall X \in \mathbb{N}: \quad \mathit{conc}(X) -\mathit{conc}(1) \; \geq \; 2 X - 2\]
instead. By choosing $\mathit{conc}(X) = 2 X$, now the resulting constraint
is valid.

So to summarize, in the first phase of our transformation,
any constraint of the form
(\ref{polycons2})
is transformed into the  unconditional constraint
\begin{equation}
\label{newpolycons}  \forall \overline{X} \in \mathbb{N}: \, \mathit{conc}(p_{n+1}) -
\mathit{conc}(q_{n+1}) \, \ge  \, \mathit{prem}(p_1,\ldots,p_n) -
\mathit{prem}(q_1,\ldots,q_n).
\end{equation}
Here, $\mathit{prem}$ and $\mathit{conc}$ are two arbitrary
new polynomials. The only requirement that we have to impose is that
 $\mathit{conc}$ must not be a constant. Indeed, if $\mathit{conc}$ would 
be a constant, then (\ref{newpolycons}) no longer implies that
(\ref{version1}) holds for all instantiations of the variables in the
polynomials $p_1,\ldots,p_{n+1},q_1,\ldots,q_{n+1}$. 
Note that we do not need a similar requirement on $\mathit{prem}$.
If a constant $\mathit{prem}$ would satisfy (\ref{newpolycons}), then 
(\ref{polycons2}) trivially holds.
 The following proposition proves the soundness of this transformation.



\begin{proposition}[Soundness of Removing Implications]\label{Soundness of First Phase of the Transformation} 
Let $\mathit{prem}$ and $\mathit{conc}$ be two polynomials with
natural coefficients, where $\mathit{conc}$ is not a constant.
Moreover, let $p_1,\ldots,p_{n+1},q_1,\ldots,q_{n+1}$ be
arbitrary polynomials with  natural coefficients.
If
\[  \forall \overline{X} \in \mathbb{N}: \;\;\; \mathit{conc}(p_{n+1}) -
\mathit{conc}(q_{n+1}) -  \mathit{prem}(p_1,\ldots,p_n) +
\mathit{prem}(q_1,\ldots,q_n) \; \geq  \; 0\]
is valid, then
\[
   \forall \overline{X} \in \mathbb{N}: \quad p_1 \ge q_1 \wedge \ldots \wedge
p_n \ge q_n \quad \Rightarrow \quad p_{n+1} \ge q_{n+1}
\]
is also valid.
\end{proposition}
\begin{proof}
For any tuple of
numbers $\overline{x}$, let
$p_i(\overline{x})$ and $q_i(\overline{x})$
denote the numbers that result from $p_i$ and $q_i$ by instantiating the
variables $\overline{X}$ by the numbers $\overline{x}$. So if $p(X_1,X_2)$ is the
polynomial $X_1^2 + 2 X_1 X_2$,
then $p(2,1) = 8$. 

Suppose that there is a tuple of numbers 
$\overline{x}$ with
$p_i(\overline{x}) \ge q_i(\overline{x})$
for all $i \in \{1, \ldots, n\}$. We have to show that then
$p_{n+1}(\overline{x}) \ge q_{n+1}(\overline{x})$ holds as well.

Since $\mathit{prem}$ only has natural coefficients, it is weakly monotonic.
Thus,
$p_i(\overline{x})\linebreak \ge q_i(\overline{x})$ for all $i \in \{1, \ldots, n\}$
implies
$\mathit{prem}(p_1(\overline{x}),\ldots,p_n(\overline{x})) \ge
\mathit{prem}(q_1(\overline{x}),\linebreak \ldots,q_n(\overline{x}))$ and thus,
$\mathit{prem}(p_1(\overline{x}),\ldots,p_n(\overline{x})) -
\mathit{prem}(q_1(\overline{x}),\ldots,q_n(\overline{x})) \ge 0$.
The prerequisites of the proposition ensure 
\[ \mathit{conc}(p_{n+1}) -
\mathit{conc}(q_{n+1}) \; \ge \; \mathit{prem}(p_1,\ldots,p_n) -
\mathit{prem}(q_1,\ldots,q_n)\]
for all instantiations of the variables. Hence, we also obtain
\linebreak
$\mathit{conc}(p_{n+1}(\overline{x})) -
\mathit{conc}(q_{n+1}(\overline{x})) \ge 0$ or, equivalently,
\begin{equation}
\label{aux} \mathit{conc}(p_{n+1}(\overline{x})) \ge
\mathit{conc}(q_{n+1}(\overline{x})).
\end{equation}
Now suppose that $p_{n+1}(\overline{x}) \not\ge q_{n+1}(\overline{x})$. Since 
$p_{n+1}(\overline{x})$ and $q_{n+1}(\overline{x})$ are \emph{numbers} (not
polynomials with variables), we would then have $p_{n+1}(\overline{x}) <
q_{n+1}(\overline{x})$. Since 
 $\mathit{conc}$ only has non-negative coefficients and since it is not a
constant, it is strictly monotonic. Thus,  $p_{n+1}(\overline{x}) <
q_{n+1}(\overline{x})$ would imply 
\[ \mathit{conc}(p_{n+1}(\overline{x})) <
\mathit{conc}(q_{n+1}(\overline{x}))\]
in contradiction to (\ref{aux}). Hence, we have $p_{n+1}(\overline{x}) \ge
q_{n+1}(\overline{x})$, as desired. 

\end{proof}

For the symbolic form of $\mathit{prem}$ and
$\mathit{conc}$, we again choose linear
or simple-mixed  polynomials. 
From our experiments, this choice
provided good results on the benchmark programs, while remaining reasonably 
efficient. 
By applying Proposition \ref{Soundness of First Phase of the Transformation},
we can now transform all constraints for the termination proof into unconditional constraints
of the form (\ref{polycons1}). If there exists a substitution of the unknown
coefficients by numbers that makes the resulting unconditional constraints valid,
then the same substitution also satisfies the original conditional constraints.

\begin{example}[applying Proposition \ref{Soundness of First Phase of the
Transformation} to the  ``\textit{der}''-program] 
\label{ex-der:sym_cons} 
We choose the decrease
condition (\ref{der:decreasecons})
in Example \ref{der:symboldecreasecond}
as an example showing how
to transform an implication 
into an unconditional constraint. 

Since the constraint (\ref{der:decreasecons}) has only one premise, here the
polynomial $\mathit{prem}$ has arity 1. We
choose a simple-mixed form for $\mathit{prem}$ and a linear form for 
$\mathit{conc}$:
%
\begin{align*}
    & \mathit{prem}(X) = \mathit{prem}_0 + \mathit{prem}_1X+\mathit{prem}_2X^2
    & \mathit{conc}(X)   = \mathit{conc}_0 + \mathit{conc}_1X.
\end{align*}
Since $\mathit{conc}$ must not be a constant, one also has to impose the constraint
            \[  \mathit{conc}_1 > 0.  \]

Now we can transform (\ref{der:decreasecons})  into an unconditional
constraint. Here, we use the following abbreviations:
\[ \begin{array}{lll}
p_1 &=&
i_0+i_1(\mathit{der}_2X^2+\mathit{der}_1X+\mathit{der}_0)+i_2\mathit{DX} \\
q_1 &=&  o_0+o_1(\mathit{der}_2X^2+\mathit{der}_1X+\mathit{der}_0)+
o_2\mathit{DX}  \\
p_2 &=& 
 d_0 +
d_1(\mathit{der}_2(\mathit{der}_2X^2+\mathit{der}_1X+\mathit{der}_0)^2+ \nonumber \\ 
&& \phantom{d_0 +
d_1(}
\mathit{der}_1(\mathit{der}_2X^2+\mathit{der}_1X+\mathit{der}_0)+\mathit{der}_0)
+d_2 \mathit{DDX} \\
q_2 &=&  d_0 +
d_1(\mathit{der}_2\mathit{DX}^2+\mathit{der}_1\mathit{DX}+\mathit{der}_0)+
d_2 \mathit{DDX} +1
\end{array}\]
Then  (\ref{der:decreasecons}) is the constraint
\[   \forall X, \mathit{DX}, \mathit{DDX} \in
\mathbb{N}:       \quad
p_1 \ge q_1 \quad \Rightarrow p_2 \ge q_2\]
and its transformation yields
\[ \begin{array}{lll}
  \forall X, \mathit{DX}, \mathit{DDX} \in \mathbb{N}:   & \mathit{conc}_0 +
\mathit{conc}_1 \, p_2 -  \mathit{conc}_0 - \mathit{conc}_1 \, q_2&\\
& -  \mathit{prem}_0 - \mathit{prem}_1 \, p_1 - \mathit{prem}_2 \, p_1^2&\\
&+ \mathit{prem}_0 + \mathit{prem}_1 \, q_1+\mathit{prem}_2 \, q_1^2 &\geq 0. \end{array}\]
By applying standard simplifications, the constraint can be rewritten to  
the following form:
\setcounter{ex:der-lastsymconsctr}{\value{equation}}
\begin{align}
       \forall X, \mathit{DX} \in \mathbb{N}: \quad &M_1X^4+M_2X^3+M_3X^2+M_4X + \nonumber\\
&  
M_5\mathit{DX}^2+M_6\mathit{DX}+
M_7 X^2 \mathit{DX} +
M_8 X \mathit{DX} +
M_9 \; \ge \;0 
 \label{ex:der-lastsymcons} 
\end{align}
where $M_1,\ldots,M_9$ are 
polynomials over the unknown coefficients
$\mathit{prem}_j$, $i_j$, $o_j$, $\mathit{der}_j$, and
$d_j$ with $j \in \{0,1,2\}$ and $\mathit{conc}_j$ with $j \in \{0,1\}$.
For example, we have
\[
M_1 \; =_{\mathit{def}} \; \mathit{conc}_1 \, d_1 \, \mathit{der}_2^3 + \mathit{prem}_2 \, o_1^2
\, \mathit{der}_2^2 -
\mathit{prem}_2 \, i_1^2 \, \mathit{der}_2^2.\]
\hfill{$\square$}
\end{example}
%
\subsubsection{Second Phase: Removing Universally Quantified Variables}\label{poly-inter-generating}\hspace*{\fill}

\vspace*{.2cm}

\noindent
In this phase, we transform any constraint of the form
\setcounter{auxctr}{\value{equation}}
\setcounter{equation}{\value{polycons1ctr}}
 \begin{equation}
   \forall \overline{X} \in \mathbb{N}: \quad p \ge 0
\end{equation}
into a set of Diophantine
constraints on the unknown coefficients. 
The transformation is again \emph{sound}:
if
there is a solution for the resulting set of Diophantine constraints,
 then 
this solution also satisfies 
the original constraint (\ref{polycons1}).
\setcounter{equation}{\value{auxctr}}

We use a straightforward transformation proposed by \cite{Hongetal98}, which
is also  used in all
related 
tools for termination of term rewriting. One only requires that all
coefficients of the polynomial $p$ are non-negative integers. Obviously, the criterion
is only sufficient, because, for instance,
$p(X) = (X - 1)^2 \ge 0$, but $X^2 - 2 X + 1$ does not have non-negative
coefficients only.

\begin{example}[removing universally quantified variables for the ``\textit{der}''-program] 
\label{positivenessexample}
We continue the transformation of Example \ref{ex-der:sym_cons}. Here, we
obtained the constraint (\ref{ex:der-lastsymcons}).
We derive the following set of Diophantine 
constraints which contains the unknown coefficients $\mathit{conc}_j$,
$\mathit{prem}_j$, $i_j$, $o_j$, $\mathit{der}_j$, and 
$d_j$ as variables: $M_1 \ge 0, M_2 \ge 0, \ldots, M_9 \ge 0$.
\hfill{$\square$}
\end{example}

\subsection{Solving Diophantine Constraints}
\label{Solving Diophantine Constraints}

The previous sections showed that one can formulate all termination conditions in
symbolic form and that one can transform them automatically into a set of Diophantine
constraints. The problem then becomes solving a system of non-linear
Diophantine constraints with the unknown coefficients as variables.
If the Diophantine constraints are solvable, then the logic program under
consideration is terminating. 
%
Solving such problems has been studied 
intensively, especially in
the context of constraint logic programming.
Moreover, there are approaches from termination of term rewriting in order to
solve such restricted Diophantine constraints automatically e.g.,
\cite{SMTCADE09,contejean05jar,Fuhsc07}.
In
\cite{Fuhsc07}, Diophantine constraints are encoded as a SAT-problem, and then a SAT
solver is used to solve the resulting SAT-problem. As shown in \cite{Fuhsc07},
this approach is significantly more efficient than solving 
Diophantine constraints by 
dedicated  solvers like \cite{contejean05jar} or by standard
implementations of constraint logic programming like in \textsf{SICStus Prolog}.

\begin{example}[solving Diophantine constraints for the``\textit{der}''-program]
\label{der:diocondsolving}
%
We start with the symbolic polynomial interpretation
from Example \ref{der:symbolrigidcond} (e.g., with
$I(\mathit{der})\linebreak = \mathit{der}_2 X^2+\mathit{der}_1 X+\mathit{der}_0$) 
and obtain the solution $\mathit{der}_2 = 1$ and $\mathit{der}_0 =
\mathit{der}_1 = 2$, 
which corresponds to
$X^2+ 2 X+2$. Similarly, we
start with the symbolic form of the polynomial interargument relation as in
Example \ref{symbolic-der}:
\begin{equation*}
    R_{d} = \{d(t_1,t_2) \mid i_0 + i_1{\mid}t_1{\mid}_I
+i_2{\mid}t_2{\mid}_I \succsim_\mathbb{N} o_0 +
o_1{\mid}t_1{\mid}_I +
o_2{\mid}t_2{\mid}_I\}.
\end{equation*}
Then we get the solution $i_1 = 1$, $i_0 = i_2 = 0$, $o_2 = 1$, $o_0 = o_1
= 0$.
This corresponds to the interargument relation
$R_{d} = \{d(t_1,t_2) \mid {\mid}t_1{\mid}_I
\succsim_\mathbb{N}
{\mid}t_2{\mid}_I\}$.
So we obtain the concrete simple-mixed polynomial interpretation from Example
\ref{exam:dist} and the concrete interargument relation from Example \ref{rigid-order:der}.
%
\hfill{$\square$}
\end{example}

\subsection{Relation to Approaches from Term Rewriting}
\label{Discussion}

Finally, we briefly discuss the connection between our approach for automated
LP termination proofs  from Section \ref{Reformulating 
the Termination Conditions Symbolically} - \ref{Solving Diophantine Constraints} and related approaches used for
termination analysis of TRSs.

Section \ref{Reformulating 
the Termination Conditions Symbolically} describes how to obtain constraints for a
symbolic polynomial order which guarantee that the requirements
of our termination criterion are fulfilled. This is similar
to related approaches used in term rewriting. Here, one also chooses
a symbolic polynomial interpretation and constructs
corresponding inequalities.
If one applies polynomial interpretations
directly for termination analysis of TRSs, then these
inequalities ensure that every rewrite rule is strictly decreasing.
If one uses more sophisticated termination techniques like the
dependency pair method \cite{ArtsGiesl00,JAR07,Hirokawa_Middeldorp04}, then one
builds inequalities which ensure 
that dependency pairs are weakly or strictly decreasing and that
rules are weakly decreasing. The decrease conditions of
dependency pairs correspond to our decrease conditions in Section \ref{Decrease Conditions}
and the requirement that rules are weakly decreasing roughly
corresponds to our symbolic constraints for valid interargument
relations in Section \ref{Valid_Interargument_Relations}. Still, there are
subtle
differences. For example, in
LPs, a predicate symbol may have several output arguments which is the
reason for the different polynomials $i_p$ and $o_p$ in our polynomial
interargument relations. Moreover, while term rewriting uses matching
for evaluation, in logic programming one uses unification. This is the
reason for our additional rigidity conditions in Section \ref{Rigidity Conditions}.

The approach in Section \ref{rewriting} shows how to find suitable values for
the symbolic coefficients. This is the same problem as in
the corresponding techniques for term rewriting. However, the
usual techniques in term rewriting can only handle unconditional
inequalities. Therefore, we have developed a new method in Section \ref{First Phase}
to remove conditions. This is a new contribution of the present paper.
In fact, after having developed this contribution for the current paper, due
to its success in the tool \textsf{Polytool}, two of the authors
of the current paper later even adapted this method to term rewriting (see
\cite[Footnote 14]{MAXPOLO}). 

The techniques of the short sections \ref{poly-inter-generating} and \ref{Solving Diophantine
Constraints}
are identical to the corresponding approaches 
used in term rewriting. We only included them here in order
to have a self-contained presentation of our approach
and to finish its illustration with the ``$\mathit{der}$''-example.

%
\section{Experimental Evaluation}
\label{sec:experiment}

In this section we discuss the experimental evaluation of our
approach. We implemented our technique in a system called \textsf{Polytool} 
\cite{Nguyen&DeSchreye06} written in \textsf{SICStus Prolog}.\footnote{For the
source code, we refer to \url{http://www.cs.kuleuven.be/~manh/polytool}.} 
Essentially, the \textsf{Polytool} system consists of four modules:
The
first module is the type inference engine,
where we use the
inference
system of 
\cite{GallagherHB05}.
The second module 
generates all termination conditions using symbolic polynomials
as in Section \ref{Reformulating 
the Termination Conditions Symbolically}. 
%
%
The third module transforms the resulting polynomial constraints 
into Diophantine constraints, as in Section
\ref{rewriting}. 
The final module is a Diophantine
constraint solver, cf.\ Section \ref{Solving Diophantine Constraints}. 
We selected the SAT-based Diophantine solver \cite{Fuhsc07} 
of the \textsf{AProVE} tool \cite{Giesletal06}.

We tested the performance of \textsf{Polytool} on 
a collection of 296 examples. The collection (Table
\ref{table1}) consists of all
benchmarks for logic programming from the \emph{Termination Problem
Data Base} (TPDB),\footnote{\url{http://www.termination-portal.org/wiki/Termination_Competition}}
where all examples that contain arithmetic or
built-in predicates were removed.  

%

{\sf Polytool} applies the following strategy:
first, we search for a linear polynomial interpretation. If
we cannot find such an interpretation satisfying the termination conditions,
then we 
search for a simple-mixed polynomial interpretation. More precisely, then we
still interpret predicate symbols by linear polynomials, but we map function
symbols to simple-mixed polynomials.
We use
  similar symbolic polynomials for 
$\mathit{conc}$ and $\mathit{prem}$ from
Section \ref{First Phase}: if the polynomial interpretation is linear, 
  then both $\mathit{conc}$ and $\mathit{prem}$ are linear. Otherwise, we use a linear
form for $\mathit{conc}$ and a simple-mixed form for $\mathit{prem}$. 
The domain for all
unknown coefficients in the generated Diophantine constraints is fixed to the set
$\{0,1,2\}$.
The experiments were performed 
on an AMD 64 bit, 2GB RAM running Linux.

We performed an experimental comparison with other leading systems for
automated termination analysis of logic programs,
namely: \textsf{Polytool-WST07}, \textsf{cTI-1.1} \cite{MesnardBagnara05}, \textsf{TerminWeb}
\cite{Codishetal99,terminWeb02}, \textsf{TALP} \cite{OhlebuschAAECC}
 and \textsf{AProVE}
\cite{Giesletal06}. 
For \textsf{TALP}, the option of non-linear polynomial
interpretations was chosen. For \textsf{cTI-1.1}, we selected the 
``default'' option. For
\textsf{AProVE} and \textsf{TerminWeb}, the fully automatic modes were
chosen. We did not include the tool \textsf{Hasta-La-Vista}
\cite{SerebrenikandDeSchreye03} in the evaluation because it is a predecessor of
\textsf{Polytool}.
We used a time limit of 60 seconds for testing each benchmark on each
termination tool. This time limit is also used in the
annual termination
  competition.

In Table \ref{table1},
we give the numbers of benchmarks which
are proved terminating (\textsf{"YES"}), the number of benchmarks which could
not be proved terminating but where processing ended within the time limit
(\textsf{"FAILURE"}), and the number of benchmarks where the tool did not stop
before the timeout (\textsf{"TIMEOUT"}). The number in square brackets is the
average runtime (in seconds) that a particular tool uses to prove termination
of benchmarks (or fails to prove termination of them within the time
limit). The detailed experiments (including also the source code of the
benchmarks and the termination proofs 
produced by the tools) can be found at 
\url{http://www.cs.kuleuven.be/~manh/polytool/POLY/journal07.html}.
Note that the two examples $\mathit{der}$ and $\mathit{div}$ presented
in this paper do not occur in the TPDB. For completeness we just mention that
\textsf{Polytool} and \textsf{AProVE} succeed on $\mathit{der}$, whereas
\textsf{cTI-1.1}
and
\textsf{TerminWeb} fail, and \textsf{TALP} reaches the timeout.
For $\mathit{div}$, all systems  except \textsf{TALP} succeed.
In the next sub-sections we discuss the results of the experiments. For a more
detailed discussion, we refer to \cite{ThangThesis}.
\begin{table}[ht]

\begin{center}
\begin{tabular}{|c|c|c|c|c|c|c|}
\hline
     & \textsf{TALP} & \textsf{cTI-1.1} & \textsf{TerminWeb}	 &  \textsf{Polytool} & \textsf{AProVE}\\ \hline
\textsf{YES} & 163 [2.54] & 167 [0.06]      & 177 [0.54]  & 214 [4.28] & 232 [6.34]             \\ \hline
\textsf{FAILURE} & 112 [1.45] & 129 [0.05]      & 118 [0.6] & 62[10.48]    & 57 [19.08]         \\\hline
\textsf{TIMEOUT} & 21  & 0      & 1   & 20     & 7            \\\hline
\end{tabular}
\end{center}
\caption{The results for 296 benchmarks of the TPDB}
\label{table1}
\end{table}

%
%
\subsection{Comparison between \textsf{Polytool} and \textsf{cTI-1.1}}
\label{polytool-cti}

Similar to \textsf{Polytool}, \textsf{cTI-1.1} deploys a global constraint-based approach to termination
analysis. However, different from \textsf{Polytool}, in \textsf{cTI-1.1} termination
inference of the analyzed program relies on its two main abstract
approximations: a program in \textsf{CLP($\mathbb{N}$)}, where all terms of
the program are mapped to expressions in $\mathbb{N}$ according to a fixed
symbolic norm (e.g., the symbolic\footnote{The difference between the
``term-size norm'' and the ``\emph{symbolic} term-size norm'' is that the
``term-size norm'' maps all variables to $0$, whereas the
``symbolic term-size norm'' maps any variable to itself (as in polynomial
interpretations).}
term-size norm by default), and a program in
\textsf{CLP($\mathbb{B}$)}, where $\mathbb{B}$ denotes the booleans, which is obtained from the
program in \textsf{CLP($\mathbb{N}$)} by mapping any number to
	$1$, any variable to itself, and addition to logical conjunction. The
		purpose of these abstractions is to capture the decrease conditions
		(the program in \textsf{CLP($\mathbb{N}$)}) and the boundedness information
			(the program in \textsf{CLP($\mathbb{B}$)}) of the program.

As shown in Table \ref{table1}, \textsf{Polytool} outperforms
\textsf{cTI-1.1}. The only benchmark where \textsf{cTI-1.1} can prove
termination and \textsf{Polytool} fails is the example
\textsf{incomplete2.pl} in the directory \textsf{SGST06} of the TPDB. 
%
%
However, if we reset the range for the values of the unknown coefficients in the
generated Diophantine constraints to $\{0,\ldots,8\}$, then \textsf{Polytool}
can prove termination for the example as well.

%
There are several reasons for the less powerful performance of \textsf{cTI-1.1}
in comparison with \textsf{Polytool}.  
First of all, \textsf{cTI-1.1} uses a fixed symbolic norm 
to map the analyzed program to a program in
\textsf{CLP($\mathbb{N}$)}, for which all termination conditions are
formulated. However, in some cases, the selected symbolic norm is not suitable
to capture the decrease in the analyzed program. Then as
a result, \textsf{cTI-1.1} cannot prove termination. The TPDB
contains a number of such benchmarks, e.g., \textsf{flat.pl},
\textsf{normal.pl} in the \textsf{talp} directory and
 \textsf{countstack.pl}, \textsf{factor.pl},
\textsf{flatten.pl} in the \textsf{SGST06} directory.

Secondly, when we use the term-size or list-length norm for
the abstract approximation in \textsf{cTI-1.1}, all constant symbols are mapped to the same
number in $\mathbb{N}$.
As a result, \textsf{cTI-1.1} fails  for examples where
the difference among constant symbols plays a role for the termination
behavior. 
In \textsf{Polytool}, different constant symbols can be mapped to different
numbers in $\mathbb{N}$. Therefore, termination of 
examples such as \textsf{simple.pl} in the
\textsf{talp} directory,
\textsf{pl2.3.1.pl} in the \textsf{plumer} directory, \textsf{at.pl} in the
\textsf{SGST06} directory, etc. can be proved, whereas \textsf{cTI-1.1} fails.

%
%

Thirdly, since termination analysis of \textsf{cTI-1.1} is based on linear
symbolic norms, it cannot prove termination of
programs such as Example \ref{exam:der} or the example \textsf{hbal\_tree.pl} in
the TPDB. 
In contrast,
\textsf{Polytool} can prove termination of these examples using simple-mixed
polynomial interpretations. 

Finally, there are examples like \textsf{applast.pl},
\textsf{bappend.pl},
\textsf{blist.pl}, \textsf{btappend.pl}, \textsf{btapplast.pl},
  \textsf{confdel.pl} and
\textsf{btree.pl} in the \textsf{SGST06} directory, 
whose termination cannot be proved by
\textsf{cTI-1.1}, since
\textsf{cTI-1.1} only uses groundness instead of type analysis.
The termination proof of these examples also fails with
 \textsf{TALP} for the same reason. In
contrast,
\textsf{Polytool} and \textsf{AProVE}
succeed for them and \textsf{TerminWeb} succeeds for some of them (i.e., \textsf{applast.pl},
	\textsf{bappend.pl}, \textsf{blist.pl}, \textsf{confdel.pl}). The
success of
  \textsf{Polytool} and \textsf{TerminWeb} is due to the use of types
  instead of modes and \textsf{AProVE} succeeds because of
so-called \emph{argument filterings} which
remove argument positions of function and
predicate symbols that are irrelevant for termination.
 But \textsf{TerminWeb} still fails on some of these examples, since it uses a fixed
  norm for part of its analysis.

A strong point of \textsf{cTI-1.1} is that it is very fast (it is by far 
the fastest tool in the experiments). The reason is
that \textsf{cTI-1.1} fixes the norm in advance. Therefore it requires much
less unknown coefficients to formulate termination conditions.  
Another strong point of \textsf{cTI-1.1} is its ability of performing
termination inference (i.e., it can try to detect all terminating modes for a
program), 
which is impossible for \textsf{Polytool} at this
moment. 
Finally, recent extensions of \textsf{cTI-1.1} include non-termination proofs,
which are not supported by the other systems in our experiments.

%
\subsection{Comparison between \textsf{Polytool} and \textsf{TerminWeb}}
\label{polytool-terminweb}

Similar to \textsf{cTI-1.1}, \textsf{TerminWeb} also uses fixed symbolic norms,
e.g., the term-size norm, the list-length norm, or (as in our experiments) a
combination of type-based norms \cite{Bruynoogheetal07} to 
approximate the analyzed program. Therefore, it has similar problems as
\textsf{cTI-1.1}. In fact, termination of examples such as \textsf{flat.pl},
\textsf{normal.pl},
 \textsf{countstack.pl}, \textsf{factor.pl},
\textsf{flatten.pl} 
discussed in Section
\ref{polytool-cti} cannot be proved by \textsf{TerminWeb} either.

Different from \textsf{Polytool} and \textsf{cTI-1.1}, \textsf{TerminWeb}
applies a local approach to termination analysis, where different norms and
level mappings
are used for different loops in the program
\cite{Codishetal99}. Hence, \textsf{TerminWeb} can prove termination of a
class of programs where lexicographic orders are required (e.g., the
benchmarks \textsf{ackermann.pl} and \textsf{vangelder.pl} in the TPDB). 
In fact, these programs could already be proven terminating by \textsf{TermiLog}
\cite{lindenstrauss97,Termilog}, the first generally available automatic
termination analyzer for LPs. \textsf{TermiLog} succeeds on these programs due to the
query-mapping pairs approach \cite{Lindenstrauss}, which has some similarity to the
dependency pair approach \cite{ArtsGiesl00,JAR07,Hirokawa_Middeldorp04}.
For termination of such programs, the global technique
based on polynomial interpretations deployed in \textsf{Polytool} is
insufficient. We are working on an extension using dependency graphs
that is able to deal with such
programs as well
\cite{Nguyenetall-LOPSTR07,LOPSTR09}.
%

Similar to \textsf{cTI-1.1}, \textsf{TerminWeb} is much faster than
\textsf{Polytool}. This is again due to the fact that \textsf{TerminWeb} uses
a fixed symbolic norm to approximate the analyzed program. 

%
%
\subsection{Comparison between \textsf{Polytool}, \textsf{AProVE}, and \textsf{TALP}}

A
point of similarity between \textsf{Polytool}, \textsf{TALP}, and \textsf{AProVE} is
that all these systems use polynomial interpretations as the  basis for the termination
analysis. 
However in \textsf{TALP} and \textsf{AProVE}, polynomial interpretations are
applied indirectly: given a 
logic program and a set of queries, these tools first transform them into a TRS
whose termination is sufficient for the termination of the original logic
program.
Then, termination analysis is applied to
the resulting TRS. Due to this transformational approach, 
several  other termination techniques developed for TRSs
become applicable for the analysis of LPs as well. In particular, \textsf{AProVE} uses
many different methods for proving
termination.

A limitation of the transformational approach in \textsf{TALP} is that it can
only handle well-moded
logic programs.
There are many non-well-moded examples in the TPDB that can
be solved by most other tools but not by \textsf{TALP}.

 \textsf{AProVE} instead applies a
quite strong transformational approach, which can also deal with non-well-moded logic
programs \cite{LOPSTR06}. Together with the powerful back-end TRS
termination prover, this makes \textsf{AProVE} a very strong LP termination system. In fact, in
both our experiments and in the termination competitions,
\textsf{AProVE} was always in the first place. In particular, it can prove
termination of most examples whenever some other tool can. Nevertheless, there exists 
one example in the TPDB (i.e., \textsf{incomplete.pl}) where
{\sf AProVE} fails to prove its termination but {\sf Polytool}
succeeds. In general, the main important observation when comparing {\sf
Polytool} and {\sf AProVE} is that although \textsf{Polytool} only
uses polynomial interpretations and {\sf AProVE} uses a large collection of
different termination techniques, {\sf Polytool} is already almost as powerful
as {\sf AProVE}. 




Similar to 
\textsf{TerminWeb}, \textsf{cTI-1.1}, and \textsf{TALP}, \textsf{AProVE} uses
mode analysis and does not provide the expressivity of types. However, it can
express classes like bounded lists, since it uses
argument filterings.
Nevertheless, in some cases,
the effect of argument filterings is not ``deep'' enough to represent
redundant argument positions adequately, cf.\ \cite{ThangThesis}.
Finally, as shown in Table \ref{table1}, \textsf{AProVE} is 
the slowest tool in the experiments. One reason is that
the transformation may generate quite complex TRSs that require more time for
termination analysis. Another reason is that \textsf{AProVE} contains much
more different termination techniques than the other tools and it tries to
apply them all after each other.

\section{Conclusions}\label{conclusion}
Since a few years, the LP  and the TRS
termination analysis communities jointly organize the ``\emph{International Workshop on
Termination}'' (WST). As a part of this workshop, the
\emph{International Competition of
Termination Tools} is organized annually,
allowing different termination tools from different categories, including
term rewriting and logic programming, to compete.
These workshops have raised a
considerable interest in 
gaining a better understanding of each other's approaches. It soon became clear
that there has to be a close relationship between one of the most popular
techniques for TRSs, polynomial interpretations, and one of the key techniques
for LPs, acceptability with linear norms and level mappings. However,
partly because of the distinction between orders over the numbers
(LPs) versus orders over polynomials (TRSs), the actual relation between the
approaches was unclear. 

One
main conclusion of the research that led to this paper is that the
distinction is a superficial one. 
%
So one outcome of our work is that,
indeed, the polynomial interpretations used for TRSs are a direct generalization of
the current practice for LPs. 

On the more technical level, the contribution of this paper is
twofold. Firstly, we provide a complete and revised theoretical framework for
polynomial interpretations in LP 
termination analysis (cf.\ Section \ref{sec:interpretation}). A first variant
of such a framework was introduced in a preliminary version of 
this paper \cite{MNTDannyd05}. Parts of this build on the results in
\cite{DeSchreyeSerebrenik01} on order-acceptability and the results in \cite{Decorteetal98}
	on the constraint-based approach for termination analysis. Another part extends the
results of Bossi et al. \cite{Bossietal91} on the syntactic characterization of
rigidity. The main revisions are in the concept of polynomial interpretations
and the concept of rigidity. Secondly, we adapt the constraint-based approach
in \cite{Decorteetal98} to represent all termination conditions symbolically, and introduce
a new approach to find such polynomial interpretations
automatically (cf.\ Section \ref{sec:automation}).

We also developed an automated tool (\textsf{Polytool}
\cite{Nguyen&DeSchreye06}) for termination proofs of LPs 
based on polynomial interpretations. 
The main contribution of the implementation
is the integration of a number of techniques including the termination
framework in Section \ref{sec:interpretation}, the call pattern inference tools in
\cite{Bruynoogheetal05,GallagherHB05,HeatonACK00,Janssensetal92},
the constraint-based approach
in Section \ref{sec:automation}, and the Diophantine constraint solver in
\cite{Fuhsc07}, to provide a completely automated termination
analyzer. 
\textsf{Polytool} participated in the annual \emph{International 
Competitions of Termination Tools} since 2007 and reached the second place, just after
\textsf{AProVE}.

We have also conducted extensive experimental evaluation for \textsf{Polytool}
and compared it empirically with other
termination analyzers such as \textsf{cTI-1.1}, \textsf{TerminWeb}, \textsf{TALP}, and
\textsf{AProVE}, cf.\ Section \ref{sec:experiment}. The evaluation shows that
\textsf{Polytool} is powerful 
enough to solve a large number of benchmarks. In particular, it can also verify
termination of examples for which non-linear norms are required. 

The current paper and the corresponding tool
provide a good basis to adapt
further techniques from the area of TRS
termination to the LP domain. In this way, 
the power of automated termination analysis can be increased
substantially. Moreover, such adaptations will clarify
the connections between the numerous termination techniques
developed for TRSs and for LPs, respectively. First steps into this direction
are
\cite{Nguyenetall-LOPSTR07,LOPSTR09}.

\section{Acknowledgements}
Manh Thang Nguyen was partly supported by \emph{GOA Inductive Knowledge Bases} and partly by
\emph{FWO Termination Analysis: Crossing Paradigm Borders}. J\"urgen Giesl and
Peter
Schneider-Kamp were supported by the \emph{Deutsche Forschungsgemeinschaft (DFG)
           grant GI 274/5-2}.
We thank John Gallagher for making his type inference engine available,
Carsten Fuhs for his SAT-based Diophantine constraint solver within
\textsf{AProVE}, Fr\'ed\'eric Mesnard and Roberto  
Bagnara for providing us the {\sf cTI} system and the \textsf{Parma Polyhedra
Library}, Michael Codish and Samir Genaim for their \textsf{TerminWeb}
system. We thank the anonymous reviewers for their valuable comments.

\bibliography{polynomial}
\end{document}